\documentclass[10pt,a4paper]{article}
\usepackage{amsthm, amsmath , amsfonts,  eucal} 
\usepackage{graphicx}

\usepackage{graphics,float}
\makeatletter
\renewcommand{\@oddhead}{\it 
Dalla Via, Fass\`o, Sansonetto: Ball in a turning cup (\today) \hfill\thepage }
\makeatother

\theoremstyle{plain}
\newtheorem{theorem}{\bf Theorem}
\newtheorem{proposition}[theorem]{\bf Proposition}

\newtheorem{definition}[theorem]{\bf Definition}

\newcommand\Sign{\mathrm{Sign}}

\newcommand{\der}[2]{\frac{\partial#1}{\partial#2}}
\newcommand{\dder}[3]{\frac{\partial^2#1}{\partial#2\partial#3}}

\newcommand\x{\mathrm{x}}
\newcommand\y{\mathrm{y}}
\newcommand\z{\mathrm{z}}

\newcommand\bList{
\begin{list}{}{\leftmargin2em\labelwidth1.5em\labelsep.5em\itemindent0em
\topsep.3ex\itemsep-.4ex} }
\newcommand\eList{\end{list}}
\newcommand\bull{\item[$\bullet$]}

\newcommand{\cC}{\mathcal{C}}
\newcommand{\cD}{\mathcal{D}}
\newcommand{\cE}{\mathcal{E}}
\newcommand{\cF}{\mathcal{F}}
\newcommand{\cG}{\mathcal{G}}

\newcommand{\cL}{\mathcal{L}}
\newcommand{\cM}{\mathcal{M}}

\newcommand{\cP}{\mathcal{P}}

\newcommand{\cR}{\mathcal{R}}

\newcommand{\bR}[1]{\mathbb{R}^{#1}}

\renewcommand{\O}[1]{\textrm{O{#1}}}
\newcommand{\SO}[1]{\textrm{SO{#1}}}

\newcommand\ugarr{\!\!\!&=&\!\!\!}

\renewcommand{\a}{\alpha}
\newcommand\g{\gamma}
\renewcommand{\o}{\omega}

\newcommand{\s}{\sigma}

\newcommand\plusarr{&&\!\!\!\!+}

\newcommand{\beq}[1]{\begin{equation}\label{#1}}
\newcommand{\eeq}{\end{equation}}

\newcommand{\bProp}[1]{\noindent\begin{Proposition}\label{#1}}
\newcommand{\eProp}{\end{Proposition}}

\newcommand{\bDef}[1]{\noindent\begin{Definition}\label{#1}}
\newcommand{\eDef}{\end{Definition}}

\newcommand{\bCor}[1]{\noindent\begin{Corollary}\label{#1}}
\newcommand{\eCor}{\end{Corollary}}

\newcommand{\bLem}[1]{\noindent\begin{Lemma}\label{#1}}
\newcommand{\eLem}{\end{Lemma}}

\newcommand{\toro}{\mathbb{T}}

\newcommand{\reali}{\mathbb{R}}

\newcommand{\rdue}{\reali^2}
\newcommand{\rtre}{\reali^3}

\newcommand\rank{\mathrm{rank\,}}

\newcommand{\SOTRE}{\mathrm{SO(3)}}

\makeatletter
\renewcommand\subsection{\@startsection{subsection}{0}{\z@}%
                                     {-3.25ex\@plus -1ex \@minus-.2ex}%
                                     {0ex}%
                                     {\normalfont\bfseries}}
\makeatother

\newcommand\D{\Delta}
\newcommand\Oj{j}
\renewcommand\O{\Omega}
\newcommand\p{\varphi}

\title{\bf On the dynamics of a heavy symmetric ball \\
that rolls without sliding \\
on a uniformly rotating surface of revolution }
\author{Marco Dalla Via\footnote{
Laboratoire Quartz EA 7393, \'{E}cole Sup\'{e}rieure d'Ing\'{e}nieurs
en G\'{e}nie \'{E}lectrique, Productique et Management
Industriel, Cergy Pontoise Cedex, France.
{\tt (E-mail: m.dallavia@ecam-epmi.com)}.
}
\footnote{
Laboratoire de Recherche en \'{E}co-innovation Industrielle et
\'{E}nerg\'{e}tique, \'{E}cole Sup\'{e}rieure d'Ing\'{e}nieurs en
G\'{e}nie  \'{E}lectrique, Productique et Management Industriel, Cergy
Pontoise Cedex, France.
} \and
Francesco Fass\`o\footnote{
Universit\`a degli Studi di Padova,
Dipartimento di Matematica ``Tullio Levi-Civita'',
via Trieste 63, 35121 Padova, Italy.
{\tt (E-mail: fasso@math.unipd.it)}.
} 
\and
Nicola Sansonetto \footnote{
Universit\`a degli Studi di Verona, 
Dipartimento di Informatica, 
Strada le Grazie 15, 37134 Verona, Italy.
{\tt (E-mail: nicola.sansonetto@univr.it)}.
}
}

\date{\small (June 30, 2021)}

\renewcommand\|{|}
\newcommand\accg{g}

\begin{document}
\maketitle
{\small
\begin{abstract}
\noindent
We study the class of nonholonomic mechanical systems formed by a
heavy symmetric ball that rolls without sliding on a surface of
revolution, which is either at rest or rotates about its (vertical)
figure axis with uniform angular velocity $\O$. The first studies of
these systems go back over a century, but a
comprehensive understanding of their dynamics is still missing.
The system has an
$\SO(3)\times\SO(2)$ symmetry and reduces to four dimensions. We
extend in various directions, particularly from the case $\O=0$ to
the case $\O\not=0$, a number of previous results and give new
results. 
In particular, we prove that the reduced system is Hamiltonizable
even if $\O\not=0$ and, exploiting the recently introduced `moving
energy', we give sufficient conditions on the profile of the surface
that ensure the periodicity of the reduced dynamics and hence  the
quasi-periodicity of the unreduced dynamics on tori of dimension up to
three. Furthermore, we determine all the equilibria of the reduced
system, which are classified in three distinct families, and
determine their stability properties. In addition to this, we give a
new form of the equations of motion of nonholonomic systems in
quasi-velocities which, at variance from the well known
Hamel equations, use any set of quasi-velocities and
explicitly contain the reaction forces.
\vskip 1truecm
\end{abstract}
}

{\small
\vskip.3cm

}

\section{Introduction}

\subsection{Motivations. }
This paper is devoted to the class of nonholonomic mechanical systems
formed by a ball that rolls without sliding on a
surface of revolution, under the action of gravity, which is assumed
to be directed as the surface figure axis. The ball is assumed to be
dynamically symmetric, namely, its center of mass coincides with its
center and its three moments of inertia relative to the center are
equal. The surface may either be at rest ($\O=0$) or rotate with
constant angular velocity $\O\not=0$ about its figure axis. 
This system has an
8-dimensional phase space, but its $\SO(3)\times\SO(2)$-symmetry
(rotate the ball about its center, and the center about the surface
figure axis) allows a reduction to dimension $4$. 

The dynamics of this system with particular---and simple---profiles of
the surface (planes, cylinders, cones) is integrable by elementary
techniques, and the first results in this direction date back at least
to the work of Routh \cite{routh}.
However, there have been relatively few general
studies of these systems, and correspondingly a global comprehension
of the dynamics with any profile is still largely missing.

When $\O=0$, the nonholonomic constraint is linear in the velocities
and the energy is conserved; being $\SO(3)\times\SO(2)$-invariant, the
energy is also a first integral of the reduced system. Routh
\cite{routh} noticed the existence of two additional $\SO(3)\times
\SO(2)$-invariant independent integrals of motion which, together with
the energy, imply that the $4$-dimensional reduced system is integrable
by quadratures. Routh also began the study of some stability
questions, mostly for $\O=0$. 

A breakthrough, in our opinion, came in the mid 1990's when the
quasi-periodicity of the system with $\O=0$ and {\it any} convex
profile was proved by Hermans \cite{hermans} and Zenkov \cite{zenkov}:
the center of the ball rotates around the figure axis and oscillates
periodically between two parallels of the surface, and the motion of
the ball about its center adds a third frequency. These results use
techniques proper to the reconstruction from periodic reduced
dynamics, see \cite{field,krupa,hermans,FGS2005}.
One of the reasons of
interest of this result is the fact that it disclosed a class of
non-Hamiltonanian integrable systems.

Another important achievement in the case $\O=0$ was, a few years
later, the discovery by Borisov, Mamaev and Kilin of the existence of a
rank-two Poisson structure in the 4-dimensional reduced space that
makes the reduced system Hamiltonian after a time-reparametrization
\cite{BMK2002}.

A non-sporadic study of the case $\O\not=0$ began in the early
2000's and lead to two main results. Borisov, Mamaev and Kilin proved
the existence of two first integrals of Routh type and of an
invariant measure of the 4-dimensional reduced system \cite{BMK2002}
(they considered the case with no gravity, but the generalizaton is
immediate). From this they deduced, via the Euler-Jacobi theorem, the
integrability by quadratures of the reduced system. 

A basic difficulty for a more detailed study of the case $\O\not=0$
was the absence of the energy integral, which is due to the fact that
if the surface rotates then the nonholonomic constraint is not linear
but affine (linear nonhomogeneous) in the velocities \cite{FS2015}.
However, two of the present authors proved that, under suitable
symmetry hypoheses, nonholonomic systems with affine constraints
possess a first integral which is a modification of the energy, and
called it a {\it moving energy} \cite{FS2016}. The existence of a moving energy
for the ball on a rotating surface was proved in \cite{FS2016}, and its
expression for this and other systems was subsequently given by
\cite{BMB2015} (who referred to it as to the `Jacobi integral'). 

Using the moving energy instead of the energy, \cite{FS2016} also proved
that the quasi-periodicity of the dynamics of the ball in a convex surface
persists if the surface rotates, at least if the angular velocity
$\O$ is sufficiently small. 

Nevertheless, at present, a general comprehension of the dynamics of
this class of systems, with any geometry of the profile, seems to be
lacking, even in the case $\O=0$. For instance important issues, such
as a general study of the equilibria of the 4-dimensional reduced
system (which are key to the comprehension of the reduced---and hence
unreduced---dynamics), has never been undertaken. Our purpose in this
paper is to begin this study, giving new results, in particular, on
its Hamiltonization, integrability and relative equilibria. 


\subsection{Content and organization of the paper. } 
We describe the system in Section 2. We limit our treatment to those
cases in which the ball rolls on a surface $\tilde\Sigma$ which is a
graph over the horizontal plane and the ball moves on top of it.
Following \cite{hermans,FGS2005}, and at variance from other
treatments \cite{routh,zenkov,BMK2002}, we assign the surface
$\Sigma$ to which the center of the ball belongs, not that on which
the ball rolls. The smoothness of $\tilde\Sigma$ puts some conditions
on the curvature of $\Sigma$, which are clarified in Proposition
\ref{p:curvaparallela}. 

The equations of motion of the system are derived in the Appendix,
as an instance of a novel form of the equations of motion of
nonholonomic systems in quasi-velocities which we derive there. At
variance from Hamel equations, that choose the
quasi-velocities so as to ``hide'' the reaction forces
\cite{hamel,NF,BMZ}, our equations use any set of quasi-velocities and
include the explicit expression of the
reaction forces as a function on the phase space (Proposition
\ref{p:PoiHam}). From a general perspective, this might be useful in
the study of a number of questions in nonholonomic mechanics in which
the reaction forces play a dominant role, such as the existence of
first integrals, invariant measures etc. 

Since the $\SO(2)$-action given by spatial rotations of the system
around the surface figure axis has isotropy, the quotient space
$M_4=M_8/\SO(3)\times\SO(2)$ is a stratified space. It consists of a
singular, one-dimensional stratum $M_4^\mathrm{sing}$ that contains
all reduced kinematical states in which the center of the ball is at
the `vertex' of the surface (the point of $\Sigma$ that belongs to
the figure axis) with zero velocity, and of a regular four-dimensional
stratum $M_4^\mathrm{reg}$. Following \cite{hermans,FGS2005} we will
embed $M_4$ in $\reali^5$ through the use of a set of 5 invariant
polynomials. This will allows us to give some results on the entire
reduced space $M_4$. Subsequently, we will specialize the analysis to
$M_4^\mathrm{reg}$ or even to its subset $M_4^\circ$ obtained by
removing all states in which the center of the ball passes (with any velocity)
through the vertex. In so doing, when this will make the description
more transparent, we will reverse to polar coordinates.

In Section 3 we study some general properties of the reduced and
unreduced systems. After giving the expressions of the two Routh
integrals and of the moving energy, extending a similar analysis in
\cite{FGS2005} we study their independence (Proposition 3). Next, we
show that the motions of the reduced system (including those that
transit through the vertex) are of four possible types (equilibria,
periodic motions, motions asymptotic to equilibria, motions which go
to infinity; Proposition 4) and we discuss their reconstruction to
the full system (Proposition 5). In particular, the already mentioned
results on the reconstruction under compact symmetry groups
\cite{krupa,field} imply that motions of the full system in relative
equilibria and relative periodic orbits are quasi-periodic on tori of
dimensions up to, respectively, two and three. 
Lastly, we prove that the level sets of the moving energy 
in $M_4$ are all compact---so that the reduced dynamics is generically
periodic and the unreduced one is generically quasi-periodic---in
two cases: if $\O=0$ and
the surface goes to $+\infty$ at infinity, and if $\O\not=0$ and the
surface goes to $+\infty$ at infinity sufficiently fast, more than
quadratically in the distance (Proposition 8). We stress that it is
only the behaviour at infinity of the surface---and no other details
of it---that plays a role in these two results. The first was in fact
proven in \cite{hermans,zenkov,FGS2005}, but was there
stated only for either convex or compact surfaces. The case $\O\not=0$ is new. (A very
weak version of it was proven in \cite{FS2016}, with a continuation
argument from the case $\O=0$, for convex surfaces and sufficiently
small $\O$'s). 

In Section 4 we restrict our analysis to the subset $M_4^\circ$ (all
states with the ball at vertex removed) and first prove the existence in
$M_4^\circ$ of a rank-two Poisson tensor that makes the system
Hamiltonian, with the moving energy as Hamilton function (Proposition
9) and the two Routh integrals as Casimirs. This tensor reduces to the
ones of \cite{ramos,FGS2005} and (up to a factor related to a time
reparameterization) of \cite{BMK2002} for $\O=0$. The interest of
this Hamiltonization result resides also in the fact that while the
Hamiltonizability of nonholonomic systems has been so far extensively
studied in the case of linear constraints, very little is known in
the case of affine constraints (the only other result we are aware of
concerns the Veselova system \cite{GN2007}). Next, we show that the
restriction of the dynamics to the level sets of the two Routh
integrals can be seen as a natural Lagrangian system with one degree
of freedom, namely with a Lagrangian which is the difference between
the kinetic energy of a point holonomically constrained to the
surface $\Sigma$ and of an `effective' potential energy which depends
on the value of the two Routh integrals (Proposition 10).

In Section 5 we determine the equilibria of the reduced system in
$M_4^\circ$, thus excluding those at the vertex (Proposition 11).  An
equilibrium of the reduced system corresponds to motions of the
unreduced system in which the center of the ball moves (or stands
still in space) on a parallel of the surface $\Sigma$, namely on a
horizontal circle, and the component of the angular velocity of the
ball normal to the surface is constant. 
We prove that there are reduced equilibria on {\it any} parallel of
$\Sigma$, which are different if the parallel is critical (a local
maximum or minimum or a saddle point of the radial height) or
regular. On each critical parallel there are two families of reduced
equilibria, the first for all $\O$'s and the second only for
$\O\not=0$, both parametrized by the vertical component
$\o_z\in\reali$ of the ball's angular velocity.
In the first family the center of the ball stands still in space;
this happens also if the surface $\Sigma$ rotates, with any $\O$. In
the reduced equilibria of the second family, instead, the center of
the ball rotates uniformly on the parallel with nonzero angular
velocity $c\O$ with a certain $0<c<1$ which depends on the moment of
inertia of the ball.
On regular parallels there is, for each $\O\in\reali$, a
family of reduced equilibria parametrized by the (nonzero)
angular velocity of the center of the ball.

In Section 6 we study the stability of the reduced equilibria,
regarding them as equilibria of the restriction of the reduced system
to a level set of the two Routh integrals, namely, to a symplectic
leaf of the rank-two Poisson structure. In order to avoid
ambiguities, we thus speak of `leafwise-stability'. This study
reduces to the study of the critical points of the effective
potential. We first give analytical conditions for the
leafwise-(in)stability of the reduced equilibria of the three
families (Proposition 12) and then we study these conditions,
with particular attention to the effect of the surface rotation.
The resulting bifurcation scenario, which is somehow rich, is
described in Propositions 13-15, and a number of situations are
considered. Overall, we reach a fairly complete understanding of the
reduced equilibria's leafwise-stability.

In Section 7 we study in some detail, and partly numerically, the
particular case in which the surface is a paraboloid. This is has two
motivations. First, since the behaviour at infinity of the surface is
exaclty quadratic in the distance from the center, our result about
the compactness of the level sets of the moving energy does not apply
when $\O\not=0$. Nevertheless, using the fact that in this case the
two Routh integrals can be explicitly determined, we can prove that
the {\it common} level sets of the three first integrals are compact,
so that the dynamics of the reduced system is generically periodic.
This suggests that our integrability results can be improved. Second,
we investigate numerically the existence and number of reduced
equilibria on the level sets of the two Routh integrals, finding that
on each of them there are between one (leafwise-stable) and three
(one of which leafwise-unstable) reduced equilibria. 

In the very short Conclusions we point out some open problems and some
future research directions.



\begin{figure}[h]
\begin{center}
{\small
{\scalebox{1.}{\includegraphics*{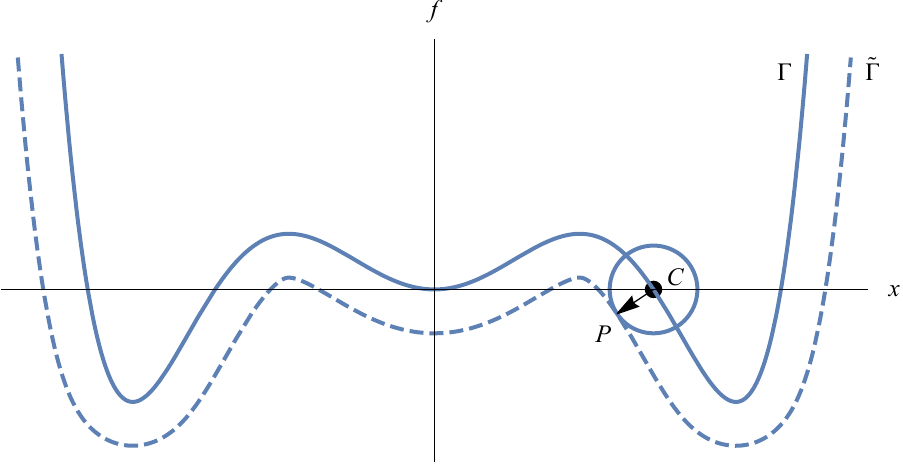}}}
}
\caption{\small The generatrices $\Gamma$ and $\tilde\Gamma$ of the
surfaces $\Sigma$ and $\tilde\Sigma$.}
\end{center}
\end{figure}


\section{The system and its reduction }
\label{s:system}

\subsection{The system. }
\label{ss:system}

We start with the holonomic system formed by a homogeneous
ball of mass $m$ and radius $a$, the center $C$ of which is
constrained to belong to a surface of revolution $\Sigma$ embedded in
$\rtre\ni(\x,\y,\z)$ and produced by the rotation, about the $\z$-axis,
of the graph $\Gamma$ of an even smooth function $f: \bR{}\to\bR{}$. More
precisely, in view of a later rescaling of the coordinates, we
assume that $\Sigma$ is described by the equation
$$
  \z = a \, f\big(\textstyle{ \frac1a \sqrt{\x^2+\y^2} }\,\big) 
  \,,\qquad (\x,\y)\in\bR2 \,.
$$
We call $f$ the `profile function' and its graph $\Gamma$ the `profile
curve'. Note that $f$ has either a minimum or a maximum at $r=0$.

The configuration manifold of this holonomic system can be identified
with $\bR2\times
SO(3)\ni(x,\cR)$, where $x=(x_1,x_2)$ are the $a$-rescaled
$(\x,\y)$-coordinates of $C$, so that $OC=(ax_1,ax_2,af(|x|))$, 
and the matrix $\cR$ fixes
the attitude of the ball. After (right) trivialization of the
tangent bundle of $SO(3)$, the phase space of the system is the
10-dimensional manifold
\[
  M_{10}
  =
  \bR{2}\times SO(3)\times \bR{2}\times\bR{3}\ni (x,\cR,\dot x,\o)\,
\]
where $\o =(\o_\x,\o_\y,\o_\z)$ is the angular velocity of the
ball relative to, and written in, the spatial frame. 

We assume that the only active force that acts on the system is
weight, directed as the downward $\z$-axis. 
We denote by $\accg$ the gravity acceleration and by $mka^2$ the
moment of inertia of the ball with respect to $C$; thus $0<k<1$
($k=\frac25$ for a homogeneous ball). Then, up to an overall
factor $ma^2$, the Lagrangian of the system is
\begin{equation}
\label{eq:Lagrangian} 
  \cL(x,\cR,\dot \cR,\o)
  =
  \frac 12 \|\dot x\|^2
  +
  \frac12 \Big( \frac {x\cdot\dot x}{\|x\|} \,f'(|x\|) \Big)^2
  +
  \frac{1}{2} k \|\o\|^2
  -
  \hat gf(|x\|) 
\end{equation}
with $\hat g=g/a$.

Next, we introduce the nonholonomic constraint that the ball rolls
without sliding on a surface $\widetilde\Sigma$ which lies {\it below}
$\Sigma$ and rotates with constant angular velocity $\Omega e_\z$
about the $\z$-axis. In the rescaled coordinates, the points of
$\tilde\Sigma$ have unit normal distance from those of $\Sigma$. The
surface $\tilde\Sigma$ is produced by the rotation of the curve
$\tilde\Gamma$ which is parallel to the graph $\Gamma$ of $f$, with
unit normal distance to it, and lies below it. It is necessary to
assume that $\tilde\Gamma$ is a regular curve and that, at each point
of contact with $\tilde\Sigma$, the ball touches $\tilde\Sigma$ in
only that point. The latter condition requires that, at each point at
which it is not concave (namely, its signed curvature is nonnegative),
the curve $\tilde\Gamma$ has radius of curvature $>1$.

As it turns out, the latter condition follows from the former, which
also ensures that $\tilde\Gamma$ is diffeomorphic to $\Gamma$:

\begin{proposition}
\label{p:curvaparallela}
$\tilde\Gamma$ is the image of a smooth immersion if and only if
\beq{curvatura}
   f''(\x) > - (1+f'(\x)^2)^{3/2} \qquad \forall \x \in\reali \,.
\eeq
In such a case, $\tilde\Gamma$ is diffeomorphic to $\Gamma$ and
has curvature radius $>1$ at each point at which it is not concave.
\end{proposition}

\begin{proof} $\Gamma$ is the image of the immersion 
$\iota:\reali\to\rdue$, $\iota(\x)=(\x,f(\x))$. The downward normal to
$\Gamma$ at the point $\iota(x)$ is
$N(\x)=\frac1{\sqrt{1+f'(\x)^2}}(f'(x),-1)$. Thus, $\tilde\Gamma$ is
the image of the map $\tilde\iota:\reali\to\rdue$ given by
$$
  \tilde\iota(\x) =
  \iota(\x)+N(\x) =
  \Big( \x + \frac{f'(\x)}{\sqrt{1+f'(\x)^2}} , 
       f(\x) - \frac1{\sqrt{1+f'(\x)^2}} \Big)\,.
$$ 
Since
$
  \tilde\iota' = 
  \big(1+\frac{f''}{(1+f'^2)^{3/2}} , 
       \big(1+\frac{f''}{(1+f'^2)^{3/2}}\big)f'\big)
$,
$\tilde\iota$ is an immersion if and only if 
$f''(x)\not=-(1+f'(\x)^2)^{3/2}$ for all $\x\in\rdue$. The fact that
$f$ is defined in all of $\reali$ rules out the possibility that
$f''(x)<-(1+f'(\x)^2)^{3/2}$ for all $\x\in\rdue$ (by a standard
comparison theorem for ODEs, since the solution of
$y'=-(1+y^2)^{3/2}$, $y(0)=0$, blows up to $-\infty$ in finite time,
if $f$ would satisfy such a condition then its derivative could not
be defined in all of $\reali$). Thus, $\tilde\iota$ is an immersion
if and only if $f$ satisfies \eqref{curvatura}.

If the signed curvature of $\Gamma$ at the point $\iota(\x)$ is
$\kappa(\x)$, then that of $\tilde\Gamma$ at the point
$\tilde\iota(\x)$ is
$\frac{\kappa(\x)}{|1+\kappa(\x)|}=:\tilde\kappa(x)$ (see e.g.
\cite{gray}). Thus, $\tilde\kappa(x)<1$ at every point $\x$ where
$\kappa(\x)>0$.

Finally, if $f$ satisfies \eqref{curvatura} then the map
$\cC:\rdue\to\rdue$,
$
\cC(\x,\z) = \big(  \x+\frac{f'(\x)}{\sqrt{1+f'(\x)^2}} \,,\,
                    \z-\frac{1}{\sqrt{1+f'(\x)^2}}\big)
$
is a diffeomorphism, and $\tilde\iota=\cC\circ\iota$.
\end{proof}

We will assume that \eqref{curvatura} is satisfied. This excludes
cases such that of a conical $\Sigma$. However, many of our results
can be applied to such cases as well after removing the vertex or
deforming the surface in a suitable neighbourhood of the vertex.
Cases in which the profile function is defined only in an open bounded
interval, and possibly diverges at its boundary, could be easily
treated as well. However we note that in such cases it might happen
that condition \eqref{curvatura} is satisifed with the opposite sign,
and this might affect the stability analysis of Section
\ref{ss:stab-RE2}.

The nonholonomic constraint forces the velocity $v_P$ of the point
$P$ of the ball in contact with the surface
$\tilde\Sigma$ to be equal to $\Omega\,e_\z\times OP$. Since $v_P=v_C +
\o\times CP$ and $OP=OC+CP$, the nonholonomic constraint is
\begin{equation}
\label{vinc1}
  v_C + \o \times CP - \Omega e_\z \times (OC+CP) =0  \,.
\end{equation}
Equation \eqref{vinc1} defines an eight-dimensional submanifold $M_8$ of
$M_{10}$ which is diffeomorphic to $\bR{2}\times SO(3)\times\bR{3}$
and can be globally parametrized with 
$(x,\cR,\dot x,\o_\z)$. Indeed, since $CP=a\,n(x)$ with\footnote{For
notational reasons, we routinely write $f$, $f'$, $F$ for
$f(\|x\|)$, $f'(\|x\|)$, $F(\|x\|)$ etc.}  
\begin{equation}
\label{n}
   n(x) := 
   \left(\frac{x_1}{\|x\|} \frac{f'}F , \frac{x_2}{\|x\|}
   \frac{f'}F ,-\frac1F \right) \,,
\end{equation}
where
\begin{equation}
\label{F}
  F \;:=\; \sqrt{ 1 +f'^2} \,,
\end{equation}
the (downward) normal unit vector to $\Sigma$ at its point
$\big(ax_1,ax_2,af\big)$, the first two entries of \eqref{vinc1} can
be written as
\begin{equation}
\label{vinc2}
  \o_\x =
  (\Omega x_1-\dot x_2)F  + (\Omega -\o_\z)\frac{x_1}  {\|x\|}f' 
  \,,\qquad
  \o_\y =
  (\Omega x_2+\dot x_1)F  + (\Omega -\o_\z)\frac{x_2} {\|x\|}f' \,.
\end{equation}
(The third equation in \eqref{vinc1} is obviously not independent of the first
two). We thus identify
$$
  M_8 = \rdue\times\SOTRE\times\rdue\times\reali 
        \ni (x,\cR,\dot x,\o_\z) \,.
$$

Clearly, the functions $\frac{ x\cdot\dot x}{|x|} f'$ and
$\frac{x_i}{\|x\|}f'$, $i=1,2$, that enter expressions 
\eqref{eq:Lagrangian} and \eqref{vinc2} are not defined at $x=0$ but
extend smoothly to $0$ at $x=0$. In order to make smoothness at $x=0$
transparent, following \cite{FGS2005} we substitute the profile
function $f$ with a smooth function $\psi:\bR{}\to\bR{}$ such that
$$
  f(r) = \psi\big( \textstyle{ \frac12{r^2}} \big) 
  \qquad
  \forall r\in\bR{} \,.
$$
The existence of such a function is granted by a result of Whitney
\cite{whitney} (see also \cite{GG}, pages 103, 108)
on account of the fact that $f$ is even. Note that
$f'(r)=r\psi'\big(\frac{r^2}2\big)$ and
$$
  \psi'\Big(\frac{r^2}2\Big) = 
  \frac{f'(r)}r 
  \qquad \mathrm{for\ } r>0  
  \,,\qquad
  \psi'(0) = f''(0) \,.
$$
However, since $f''(r)=\psi'\big(\frac{r^2}2\big) 
+r^2\psi''\big(\frac{r^2}2\big)$ and
$$
 \psi''\Big(\frac{r^2}2\Big) 
  =
  \frac{rf''(r)-f'(r)}{r^3}
 \quad \mathrm{for\ } r>0  \,,
$$
we will use $f''$ when we need to stress the dependence on the
convexity properties of the profile.

The equations of motion of this nonholonomic system are derived in the
Appendix. We need them only as a tool to deduce those of the reduced
system.

\subsection{The $\mathbf{SO(3)\times SO(2)}$-reduced system. }
\label{ss:reduction}

Consider now the action $\Xi$ of $SO(3)\times SO(2)$ on $M_{10}$
given by
$$
   \Xi_{(S,P)}(x,\cR,\dot x,\o) = (Px,P\cR S,P\dot x,P\o) \,,
$$
namely, $SO(3)$ acts on the right on itself and
$SO(2)$ acts by rotations about the $z$ axis. From \eqref{vinc2} it
follows that the constraint manifold $M_8$ is invariant under the action
$\Xi$. Therefore, $\Xi$ restricts to an action on $M_8$. Moreover,
since the Lagrangian \eqref{eq:Lagrangian} is invariant under
$\Xi$, the equations of motion of the nonholonomic system in
$M_8$ are invariant under the restriction of $\Xi$ to $M_8$
\cite{BS,BKMM} and can be reduced to $M_8/(SO(3)\times SO(2))$. Since
the actions of $SO(3)$ and $SO(2)$ commute, the reduction can be
performed in stages. 

Since the Lagrangian and the constraint are independent of the
attitude $\cR$ of the ball, the $SO(3)$-reduction consists in simply 
cutting off the $SO(3)$ factor of $M_8$, and the $SO(3)$-reduced
space is the five-dimensional manifold
\[
  M_5 = \bR{2}\times\bR{2}\times\bR{}\ni(x,\dot x,\o_\z) \,.
\]
The $SO(2)$-action on $M_{8}$ induces an action on $M_5$ given by
$P.(x,\dot x,\o_\z)= (Px,P\dot x,\o_\z)$, which is free at all points
of $M_5$ except at those with $x = \dot x = 0$ (the kinematical
states in which the ball is at the vertex of the surface and the
velocity of its center of mass is zero---hence, its angular velocity
is vertical).\footnote{The invariance of the singular stratum
$M_4^\mathrm{sing}$, see below, implies that these kinematical states
are met only in motions in which the ball stands at the vertex
spinning around the vertical axis.} 

The reduction under this action is well known. In fact, $SO(2)$ does
not act on the $\bR{}$-factor of $M_5$, while its action on the factor
$\bR{2}\times\bR{2}$ is nothing but the familiar $SO(2)$-action of
the 1:1 oscillator \cite{hermans,FGS2005}. Therefore, the reduced
space $M_5/SO(2)=M_8/SO(3)\times SO(2)$ can be identified with the
semialgebraic variety 
\[
  M_4
  = 
  \big\{ (p_0,p_1,p_2,p_3,p_4) \in\bR{5} \;:\; 
  4p_0 p_1 =p_2^2+p_3^2 \,,\;  p_0 \ge0\,,\; p_1 \ge0 \big\} 
\]
immersed in $\bR5 \ni (p_0,p_1,p_2,p_3,p_4)=:p$,
with quotient map $M_5\to M_4$
given by
\begin{equation}
\label{PolInv}
  p_0  = \frac{1}{2}\, \|\dot x\|^2 
  \,,\quad 
  p_1 =\frac{1}{2}\,\|x\|^2
  \,,\quad
  p_2 = x\cdot \dot x
  \,,\quad 
  p_3 = x_1\dot x_2 - x_2\dot x_1
  \,,\quad 
  p_4 = \o\cdot n(x)
\end{equation}
(a set of generators of the invariant polynomials of the
$SO(2)$-action, see \cite{hermans,CDS}; see also \cite{BMK2002}).

The last coordinate $p_4$ for $\bR5$ has been chosen as $\o\cdot
n$, instead of $\o_\z$, because this will somehow simplify the expression,
and the analysis, of the moving energy. It also simplifies the
equations that define the other two first integrals of the system,
$J_1$ and $J_2$ below, but this is actually not that important. 

The semialgebraic variety $M_4$ consists of two strata: a
``singular'' one-dimensional stratum
$$
  M_4^\mathrm{sing}= \{p\in\bR{5}\; : \; p_0  = p_1 =p_2=p_3 = 0\} 
  \approx \bR{}\ni p_4\,,
$$
which is the quotient of the one-dimensional submanifold
$M_5^\mathrm{sing}= \{(0,0)\}\times\{(0,0)\}\times\bR{}$ of $M_5$ left fixed
by the $SO(2)$-action, and can be identified with it, and a
four-dimensional ``regular'' stratum 
$$
   M_4^\mathrm{reg} =
   \big\{p\in\bR{5} \;:\; 4p_0 p_1 =p_2^2+p_3^2\,,\; 
   p_0 \ge 0\,,\; p_1 \ge 0\,,\; (p_0 ,p_1 )\not=(0,0) \big\}
   \,,
$$
which is the quotient of the open subset of $M_5$ 
where the $\SOTRE\times\SO(2)$-action is free. 

We will denote
$$
   \pi:M_8\to M_4
$$
the quotient map associated to the $\SOTRE\times\SO(2)$-action in
$M_8$. Note that then
$$
   M_4^\mathrm{reg}=\pi(M_8^\mathrm{reg})
$$
with $M_8^\mathrm{reg} = (\rdue\setminus\{0\}) \times\SOTRE\times
(\rdue\setminus\{0\}) \times \reali$.

At a certain stage we will restrict to the submanifold of $M_4^\mathrm{reg}$
where $p_1>0$, which is diffeomorphic to $\reali_+\times\rtre$ and can
be globally parametrized with either $(p_1,p_2,p_3,p_4)$ or $(r,\dot
r,\dot \theta,\o_n)$ (or, for that matter, with $(r,\dot r,\dot
\theta,\o_\z)$ as well).
In fact, we will switch between these two parametrizations depending
on the needs: the former is closely linked to the theory in $M_4$ and
$M_4^\mathrm{reg}$, the latter has a more direct physical interpretation.

\vskip4mm
{\it Remark: } The manifold $M_4^\mathrm{reg}$ is diffeomorphic to
$(\bR3\setminus\{0\})\times\bR{}$, with global parametrization
$
  (\rtre\setminus\{0\})\times\reali
  \ni \big((y,p_2,p_3),p_4\big)
  \,\mapsto\,
  \Big(\frac12\Big(\sqrt{y^2+p_2^2+p_3^2}\,-\,y\Big), 
       \frac12\Big(\sqrt{y^2+p_2^2+p_3^2}\,+\,y\Big), 
       p_2,p_3,p_4 \Big) \,.
$ 
However, we will prefer using its embedding in $\reali^5$. 

\subsection{The equations of motion of the reduced system. }
\label{ss:RedVF}

Following \cite{hermans,FGS2005}, we write the equations of motion of
the $\SOTRE\times\SO(2)$-reduced system in $M_4$ (from now on,
`reduced system') as the restriction to $M_4$ of a set of
equations in $\bR5$. The deduction of these equations is done in the
Appendix, on the basis of a new form of the equations of motion of
nonholonomic systems.

The equations of motion of the reduced system are the
restriction to $M_4$ of the equation
\begin{equation}
\label{EqRid}
  \dot p = X(p)  \,,\qquad p=(p_0 ,p_1,p_2,p_3,p_4) \in\bR5 \,
\end{equation}
where $X=(X_0,X_1,X_2,X_3,X_4)$ is the vector field
in $\bR5$ with components
\begin{equation}
\label{X}
\begin{aligned}
  X_0& = 
    p_2\Big( \big(\mu p_3 p_4\psi'' -p_2^2\psi'\psi''
                    - \gamma \psi' -2p_0 \psi'^2\big)\cF^2
   + \Omega \mu p_3\big(\psi'^2+\cF\psi'') \cF^2 \Big)
  \\  
  X_1& = p_2 
  \\  
  X_2& =
    \big( 2p_0 -\mu p_3 p_4\psi' -2 \g p_1 \psi'-
                    2p_1 p_2^2 \psi'\psi''\big)\cF^2 
    - \Omega \mu p_3 \big(1+\psi'\cF\big)\cF^2 
  \\
  X_3 & = p_2\, \big( G_3 p_4+ \Omega g_3\big)
  \\
  X_4 & = p_2\, \big(G_4 p_3 + \Omega g_4\big)
\end{aligned}
\end{equation}
where
\begin{equation}
\label{simboli}
  \mu= \frac k{1+k} \,,\qquad  \g = \frac{\hat g}{1+k} \,,\qquad
  \cF(p_1)=\frac1{F(\sqrt{2p_1})} =\frac1{\sqrt{1+2p_1\psi'(p_1)^2}} \,,
\end{equation}
and
\begin{equation}
\label{Gg}
\begin{aligned}
  G_3& = \mu \big(\psi' + 2 p_1  \psi''\big) \cF^2 \,,&\qquad
  g_3& = \mu \big( 1+(\psi' + 2 p_1 \psi'')\cF^3  \big)
  \\
  G_4& = \big(\psi'^3 - \psi''\big) \cF^2 \,,&
  g_4& = \big(1+\cF\psi'\big)\, \big( \psi'+2p_1 \psi''\big)\cF^2
  \,.
\end{aligned}
\end{equation}
Note that $\frac12<\mu<1$ and that $\psi$, $\cF$, $G_3$,
$G_4$, $g_3$ and $g_4$ are functions of $p_1$ alone
and are independent of $\Omega$. Instead, $f$ and $F$ are functions of
$r$, and $F(r)=1/\cF(r^2/2)$.

For consistency, we note that $M_4$ is invariant under the flow of the
vector field $X$ in $\bR5$: $X$ vanishes at the points of $M_4^\mathrm{sing}$ and
is tangent to $M_4^\mathrm{reg}$ given that $L_X(p_2^2+p_3^2-4p_0 p_1 )=0$. 

From \eqref{X} it follows that the equilibria of the reduced system
are the points where $p_2=0$ and $X_2=0$. They are all the points of
the singular stratum $M_4^\mathrm{sing}$ and the points of the set
\begin{equation}
\label{cE}
  \mathcal{E}_4^\mathrm{reg}
  =
  \big\{ p\in M_4^\mathrm{reg} \,:\; p_2 = 0\,,\; 
  X_2(p_0 ,p_1,0,p_3,p_4) = 0 \big\} \,.
\end{equation}
The reduced equilibria forming the singular stratum
$M_4^\mathrm{sing}$ are the projection of relative equilibria in
$M_8$ which consist of motions in which the ball stands at the vertex
of the surface and uniformly spins with constant, vertical angular
velocity. Relative equilibria that project onto reduced equilibria in
$\mathcal{E}_4^\mathrm{reg}$ consist instead of motions of the
nonholonomic system in $M_8$ in which the ball uniformly rolls along
a horizontal circle in $\tilde \Sigma$. We will study reduced
equilibria in $\mathcal{E}_4^\mathrm{reg}$ and their stability in 
Section~\ref{s:RedEq}. Instead, we will not study in this work the
stability of the reduced equilibria in $M_4^\mathrm{sing}$, and the
related existence of motions asymptotic to/from them, because that
would require the analysis of the system in the $\SOTRE$-reduced space
$M_5$, which is extraneous to the approach taken here and is left
for a separate work.

Finally, we note that the dynamics of the reduced system relative to
a certain $\O\not=0$ is conjugate by the reflection
\beq{C}
  C:M_4^\mathrm{reg}\to M_4^\mathrm{reg}\,, \qquad
  C(p_0,p_1,p_2,p_3,p_4)=(p_0,p_1,p_2,-p_3,-p_4) 
\eeq
to that of the reduced system relative to $-\O$. In fact, if we make
momentarily explicit the dependence of the vector field $X$ on the
surface's angular velocity $\Omega$ by denoting it $X_\Omega$, it
follows from \eqref{X} that
\beq{ReflSimm}
   C_*X_\Omega = X_{-\Omega} \qquad \forall \O\in\reali \,.
\eeq
In particular, the dynamics at $\O=0$ is invariant under the
reflection $C$.

\section{Reduced and unreduced dynamics}
\label{s:Int}

In this Section we first describe some general features of the dynamics of
the reduced and unreduced systems and then particularize to the case
of coercive profile functions.

\subsection{The first integrals. }
\label{ss:FI}
The reduced system (and hence the unreduced one) is known to have
three integrals of motion: the moving energy discovered in
\cite{FS2015} and two other integrals, whose existence was proven
by Routh for $\O=0$ (and for the special case
of a spherical profile also for $\O\not=0$, \cite{routh}, section
224) and by Borisov, Mamaev and Kilin for $\O\not=0$
\cite{BMK2002}.   
In order to express the latter two integrals
we note that the equations for $p_3$ and $p_4$ are
\begin{equation}
\label{p3p4}
   \Big( \begin{matrix} \dot p_3 \\ \dot p_4 \end{matrix} \Big)
   = p_2 \Big[ G(p_1 )\Big(\begin{matrix} p_3 \\ p_4 \end{matrix} \Big) + \Omega
   g(p_1 ) \Big] 
\end{equation}
where
\begin{equation}
\label{Gg-2}
  G(x) :=\left( \begin{matrix} 0 & G_3(x) \\ G_4(x) & 0\end{matrix} \right)
  \,,\qquad 
  g(x) := \left( \begin{matrix} g_3(x) \\ g_4(x) \end{matrix} \right) 
\end{equation}
with $G_3$, $G_4$, $g_3$ and $g_4$ as in \eqref{Gg}.
Let $\bR{}\ni x\mapsto U(x)\in GL(2)$ be the solution of the
matrix differential equation
\begin{equation}
\label{naode1}
  U' = G(x)U \,, \qquad U(0)=\mathbb I 
  \qquad \big(U\in GL(2) \big)
\end{equation}
and $\bR{}\ni x\mapsto u(x)\in\bR2$ the solution of the differential
equation
\begin{equation}
\label{naode2}
   u'=G(x)u+g(x) \,,\qquad u(0)=0 \qquad \big(u \in \bR2\big)
\end{equation}
(recall that linear (non)homogeneous
equations have global existence of the solutions).

\begin{proposition}
\label{p:FI} The restrictions to $M_4$ of the function $E:\bR5\to\bR{}$
given by
\begin{equation}
\label{E}
  E \;=\; \g\psi + p_0  + \frac12 p_2^2 \psi'^2 +
    \frac12\mu p_4^2
    \;+\;
  \Omega\big(\mu p_4 \cF -p_3\big)
  \;+\; \Omega^2 \mu p_1  \big(1-\cF^2 \psi'^2 \big) 
\end{equation}
and of the two components $J_1, J_2$ of the map $J:\bR5\to\bR2$ given by
\begin{equation}
\label{J}
  J = U(p_1 )^{-1} \Big[
     \Big( \begin{matrix} p_3 \\ p_4 \end{matrix} \Big)
    -\Omega \,u(p_1 ) \Big] 
\end{equation}
are first integrals of the reduced system \eqref{EqRid}.
\end{proposition}

\begin{proof} We show that $E,J_1,J_2$ are first integrals of system
\eqref{EqRid} in the entire $\bR5$. That $L_XE=0$ is checked with a
computation. If we denote with a dot the derivative with respect to
time and with a prime the  derivative with respect to $p_1 $, then,
along a solution of \eqref{EqRid}
$$
  \frac d{dt}J =
  \dot p_1  
  (U^{-1})'
      \Big(
         \Big( \begin{matrix} p_3 \\ p_4 \end{matrix} \Big)
         - \Omega u )
      \Big) 
  +
  U^{-1}
     \Big(
        \Big( \begin{matrix} \dot p_3 \\ \dot p_4 \end{matrix} \Big)
        - \Omega \dot p_1  u'
     \Big) \,.
$$ 
The fundamental matrix $U$ satisfies the equation $U'=GU$, which
implies $(U^{-1})'=-U^{-1}G$. Using this equality, $u'=Gu+g$ and
\eqref{p3p4} one verifies that $\frac d{dt}J
=0$.\end{proof}

We will refer to $E|_{M_4}$ as to the `reduced moving energy' of
and to $J_1|_{M_4}$ and $J_2|_{M_4}$ as to `reduced Routh
integrals' of the system. The pull-backs of these functions to $M_8$
give three $\SOTRE\times\SO(2)$-invariant first integrals of the
unreduced system. 

We also note that, if we momentarily make explicit the dependence of the first
integrals on $\Omega$ by denoting them $E_\O$ and
$J_\O=(J_{\O,1},J_{\O,2})$, then
\begin{equation}
\label{ReflRouth}
   E_\O = E_\O\circ C \,,\qquad
   J_\O \circ C=-J_{-\O} \qquad \forall \O\in\reali 
\end{equation}
where $C$ is the reflection \eqref{C}.

We now prove that the three first integrals are everywhere
functionally independent at all points of $M_4$ which are not
equilibria. Specifically, we neglect the singular stratum
$M_4^\mathrm{sing}$ (which consists of equilibria) and prove that
$E,J_1,J_2$ are functionally independent at all points of the regular
stratum $M_4^\mathrm{reg}$ but the equilibria. For $\Omega=0$ this was
proven in \cite{FGS2005} with a direct computation. For
$\Omega\not=0$ a direct computations is somewhat cumbersome and we
use a somehow different argument. This argument makes explicitly
appear in the proof the component $X_2$ of the reduced vector field
and in this way sheds some light on why, in $M_4^\mathrm{reg}$, the
independence is lost exactly at the reduced equilibria. 

Let us define two functions $\tilde p_3,\tilde
p_4:\reali\times\rdue\to\reali$ as
\beq{tildep3p4}
  \left(\begin{matrix} \tilde p_3(p_1,j) \cr \tilde p_4(p_1,j) \end{matrix}\right) 
  \,:=\,
  U(p_1)j+\Omega\, u(p_1) 
\eeq
with $U$ and $u$ as in Proposition \ref{p:FI}. 

\begin{proposition}
\label{p:submersion}
\ 
\bList
\item[i.] The critical points of the map
$(E,J)|_{M_4^\mathrm{reg}}:M_4^\mathrm{reg}\to\bR3$ are the points of the set
$\mathcal{E}_4^\mathrm{reg}$.
\item[ii.] The map $J|_{M_4^\mathrm{reg}}:M_4^\mathrm{reg}\to\bR2$ is a
surjective submersion. 
\eList
\end{proposition}

\begin{proof}
(i.) $M_4^\mathrm{reg}\subset \bR5$ is one of the two components of the zero
level set of the function $K:\bR5\to\bR{}$,
$K(p)=\frac{p_2^2+p_3^2}{2} - 2 p_0 p_1$, with the singular
stratum $M_4^\mathrm{sing}=\{(0,0,0,0)\}\times \bR{}$ removed. We determine the
critical points of $(E,J_1,J_2)|_{M_4^\mathrm{reg}}$ at the points of
$K^{-1}(0)$ using Lagrange multipliers
$\lambda=(\lambda_1,\lambda_2,\lambda_3,\lambda_4)$. The critical 
points of $(E,J_1,J_2)|_{M_4^\mathrm{reg}}$ in $K^{-1}(0)$ are those at
which the equation
\begin{equation}
\label{LagrMult}
 \lambda_1 dJ_1 + \lambda_2 dJ_2 + \lambda_3 dE + \lambda_4 dK = 0
\end{equation}
has a nontrivial solution $\lambda\not=0$.
For notational convenience we introduce the function $\cG_\lambda :=
\lambda_1J_1 + \lambda_2J_2 + \lambda_3E + \lambda_4K:\bR5\to\bR{}$,
where the $\lambda_i$'s have to be thought of as parameters
(namely, $d\cG_\lambda$ equals the left hand side of \eqref{LagrMult}).

We begin noticing that $\partial_{p_0 }\cG_\lambda =
\lambda_3 - 2p_1 \lambda_4$ and $\partial_{p_2}\cG_\lambda=
p_2\psi'^2 \lambda_3+ p_2\lambda_4$ vanish simultaneously in the
following three\footnote{Not two, as erroneously stated, in the case
$\Omega=0$, in \cite{FGS2005}.} cases: (a) $\lambda_3 = \lambda_4=0$,
(b) $\lambda_3=0$, $\lambda_4\not=0$, $p_1 =p_2=0$, (c) $\lambda_3 =
2p_1 \lambda_4$, $p_2=0$, $p_1 \not=0$, $\lambda_4\not=0$.

The first two cases do not lead to any critical point in $M_4^\mathrm{reg}$. In
case (a), \eqref{LagrMult} reduces to $\lambda_1dJ_1+\lambda_2dJ_2=
0$ and hence admits only the trivial solution because the two
functions $J_1,J_2:\bR5\to\bR{}$ are functionally independent given
that the fundamental matrix $U$ is nonsingular. In case (b), since
$\lambda_3=0$, $\partial_{p_1 }\cG_\lambda|_{p_1 =p_2=0} =
-2p_0 \lambda_4$ which, for $\lambda_4\not=0$, vanishes only if
$p_0 =0$: but there are no points in $M_4^\mathrm{reg}$ with $p_0 =p_1 =0$.

We thus consider case (c). We may assume $\lambda_4=1$,
$\lambda_3=2p_1 $. The vanishing of
$\partial_{p_3}\cG_\lambda\big|_{p_2=0}$ and
$\partial_{p_4}\cG_\lambda|_{p_2=0}$ gives the linear system for
$\lambda_1,\lambda_2$
$$
  (DJ)^T \Big(\begin{matrix} \lambda_1\\\lambda_2\end{matrix}\Big) 
  = 
  -
\nabla_{(p_3,p_4)}\big(\lambda_3E+K)\big|_{p_2=0,\lambda_3=2p_1 } 
$$
where $DJ$ stands for the Jacobian matrix of $J=(J_1,J_2)$ with
respect to $(p_3,p_4)$. Since $DJ=U^{-1}$ is nonsingular,
this system determines the multipliers $\lambda_1,\lambda_2$:
$(\lambda_1,\lambda_2)=\ell$ with
$$
 \ell = -
 U^T\nabla_{(p_3,p_4)}\big(\lambda_3E+K)\big|_{p_2=0,\lambda_3=2p_1 } 
 \,.
$$
Thus, equation \eqref{LagrMult} reduces to the only condition
$
\partial_{p_1 }
\cG_\lambda\big|_{p_2=0,\,\lambda=(\ell_1,\ell_2,2p_1 ,1)} =0 
$,
namely
\begin{equation}
\label{last}
  \ell\cdot \partial_{p_1 }J +
  \partial_{p_1 } (\lambda_3E+K)\big|_{p_2=0,\,\lambda_3=2p_1 } = 0 \,.
\end{equation}
Let us shorten $(p_3,p_4)=:y$, denote by a prime the
derivative with respect to $p_1 $ and write $J'$ for $(J'_1,J_2')$.
From \eqref{J}, $J'=(U^{-1})'(y-\Omega u) - \Omega U^{-1} u'$. As
already noticed, $(U^{-1})'=-U^{-1}G$ and $u'=Gu+g$. Thus
$J'=-U^{-1}(Gy+\Omega g)$ and so $\ell\cdot J'= (Gy+\Omega g)\cdot
\nabla_{(p_3,p_4)}\big(\lambda_3E+K)\big|_{p_2=0,\lambda_3=2p_1
}$. Therefore, condition \eqref{last} is
\begin{equation}
\label{last2}
  \big[ (Gy+\Omega g) \cdot \nabla_{(p_3,p_4)}+ \partial_{p_1 }\big]
  (\lambda_3E+K)\Big|_{p_2=0,\lambda_3=2p_1 } 
  =0 \,.
\end{equation}
Note now that, since $E$ and $K$ are first integrals of system
\eqref{X} in $\bR5$, $L_X(\lambda_3 E+K)=0$ and therefore, for all
$p_2$ and $\lambda_3$,
$$
  p_2\big[ (Gy+\Omega g) \cdot \nabla_{(p_3,p_4)}+
\partial_{p_1 }\big]
  (\lambda_3E+K)  
  =
  - \big( X_0\partial_{p_0 } + X_2\partial_{p_2} \big) (\lambda_3E+K)  
$$
Hence, for all $p_2\not=0$ and all $\lambda_3$,
$$
  \big[ (Gy+\Omega g) \cdot \nabla_{(p_3,p_4)}+ \partial_{p_1 }\big]
  (\lambda_3E+K)
  =
  - \frac1{p_2} \big( X_0 \partial_{p_0 } + X_2\partial_{p_2} \big)
  (\lambda_3E+K)  \,.
$$
But $\partial_{p_0 }(\lambda_3E+K) =\lambda_3-2p_1 $ vanishes for
$\lambda_3=2p_1 $ while $\partial_{p_2}(\lambda_3E+K) =p_2(1+\psi'^2)$.
Hence, for $p_2\not=0$,
$$
  \big[ (Gy+\Omega g) \cdot \nabla_{(p_3,p_4)}+ \partial_{p_1 }\big]
  (\lambda_3E+K)\big|_{\lambda_3=2p_1 } 
  =
  -(1+\psi'^2) X_2
  \,.
$$
By continuity,
this equality is satisfied at $p_2=0$ as well. Hence, \eqref{last2} is
equivalent to $p_2=0$, $X_2|_{p_2=0}=0$, which defines the zeroes of
$X$ in $M_4^\mathrm{reg}$, see \eqref{cE}. 

(ii.) Surjectivity of $J|_{M_4^\mathrm{reg}}:M_4^\mathrm{reg}\to\rdue$ is obvious. 
In order to verify that it is a submersion, put $\lambda_3=0$ in the
previous computations. The vanishing of $\partial_{p_0
}\cG_\lambda = - 2p_1 \lambda_4$ and $\partial_{p_2}\cG_\lambda=
p_2\lambda_4$ gives either $\lambda_4=0$ (hence, as before,
$\lambda_1=\lambda_2=0$) or $p_1 =p_2=0$ (which is not satisfied
at any point in $M_4^\mathrm{reg}$).
\end{proof}

\vskip4mm
{\it Remarks. } 
(i) The pull-back of $E|_{M_4}$ differs by a factor $k+1$ from
the reduced moving energy of the (unreduced) system as defined in
\cite{FS2015}. The existence of this first integral was proven in
\cite{FS2015} and its expression was then computed in \cite{BMB2015}. 

(ii) With reference to the theory developed in
\cite{FS2015,FNS2018}, we note that the reduced moving energy of the
(unreduced) system is the difference between the energy
$E_0=\cL+2\hat g f$ and the `momentum' of the vector field 
$Y=\big(-\O\frac{x_2}{|x|},\O\frac{x_1}{|x|},0,0,\Omega\big)$ on
the configuration manifold $\bR2\times SO(3)$ of the system. This is
a `kinematically interpretable' moving energy in the sense of
\cite{FNS2018} and its conservation follows from Proposition 8 of
\cite{FNS2018}.

(iii) As shown in \cite{FGS2009}, when $\Omega=0$ the Routh integrals
are ``gauge momenta'' \cite{FGS2008}. In the case of the
rotating cylinder the two Routh integrals are gauge momenta as well
\cite{FS2015}. 
In analogy with the case of linear constraints \cite{FGS2012}, the fact that,
being $\SOTRE\times S^1$-invariant, the Routh integrals are
``weakly-Noetherian'' (in the sense of \cite{FGS2008}) might suggest that
they are always gauge momenta.

\subsection{Some results on the reduced and unreduced dynamics. }
\label{ss:Dyn}

The existence of three independent integrals of motion makes the
reduced dynamics in $M_4$ very simple. 

\begin{proposition}
\label{p:RedDyn}
Assume that $p\in M_4$ is not an equilibrium point of $X$ and let
$\eta_p$ be the connected component of the fiber of $(E,J)|_{M_4}$ that
contains $p$.
\bList
\item[i.] If $\eta_p$ does not contain any equilibrium, then the
integral curve of $X$ through $p$ either is periodic or leaves any
compact subset of $M_4$ for both positive and negative times.
\item[ii.] If $\eta_p$ contains an equilibrium, then for positive
times the integral curve of $X$ through $p$ either leaves any compact
subset of $M_4$ or is asymptotic to an equilibrium. The same happens
for negative times.
\eList
\end{proposition}

\begin{proof}
(i.) Not containing equilibria, $\eta_p$ is a subset of
$M_4^\mathrm{reg}\setminus\mathcal{E}_4^\mathrm{reg}$ and, by
Proposition \ref{p:submersion}, is a component of a regular fiber of
$(E,J)|_{M_4}$. As such, $\eta_p$ is a closed embedded one-dimensional
submanifold of $M_4^\mathrm{reg}$, which is moreover invariant under
the flow of $X$ and does not contain any equilibrium. Thus, $\eta_p$
is the image of the maximal integral curve of $X$ through $p$.  If
$\eta_p$ is diffeomorphic to $S^1$, then the integral curve of $X$
through $p$ is periodic. If $\eta_p$ is diffeomorphic to $\reali$,
then it is parametrized by the maximal integral curve of $X$ through
$p$, say $\p:(T_-,T_+)\to M_4$ with $\p(0)=p$ and some $-\infty\le
T_-<0<T_+\le+\infty$. Assume now, by contradiction, that
$\eta_+:=\p([0,T_+))$ is contained in a compact subset $K$ of $M_4$.
Then $T_+=+\infty$ and, since $\eta_p$ is an embedded submanifold,
$\lim_{t\to+\infty}\p(t)=:p_+$ exists in $K$. Elementary facts about
ODEs imply that then $X(p_+)=0$. But this is impossible because
$p_+\in \eta_p$, given that $\eta_p$ is closed, and $\eta_p$ does not
contain equilibria. Similarly for $\eta_-:=\p((T_-,0])$. 

(ii.) Let $\eta^\mathrm{eq}$ be the set of points of $\eta_p$ at
which $X$ vanishes. Thus $\eta^\mathrm{eq} =
\eta_p\cap(M_4^\mathrm{sing}\cup\mathcal{E}_4^\mathrm{reg})$ and
$\eta_p\setminus\eta^\mathrm{eq} \subset M_4^\mathrm{reg}
\setminus \cE_4^\mathrm{reg}$. Let $\eta_p^*$ be the connected
component of $\eta_p\setminus\eta^\mathrm{eq}$ that contains $p$.
$\eta_p^*$ is $X$-invariant and is a connected component of a fiber of 
$(E,J)|_{M_4^\mathrm{reg} \setminus \cE_4^\mathrm{reg}}$. Since
$M_4^\mathrm{reg} \setminus \cE_4^\mathrm{reg}$ is an open subset of
$M_4^\mathrm{reg}$, $\eta_p^*$ is a one-dimensional immersed
submanifold of $M_4$. Being $X$-invariant, $\eta_p^*$ is the image of
the maximal integral curve of $X$ through $p$. At variance from case i.,
however, now $\eta_p^*$ is not closed. Thus, the integral curve
through $p$ either leaves every compact set or tends to an
equilibrium point. \end{proof}

We note that reduced motions may leave any compact set in $M_4$ in
two ways: either the center of the ball goes to infinity or some
components of the velocity go to infinity. The conservation of the
moving energy, together with the `Hamiltonization' of the reduced
system which shows that it is a family of one-degree-of-freedom
Hamiltonian (or Lagrangian) systems of mechanical type, 
(Proposition \ref{p:Lagr}) will imply
that the latter possibility can only take place with motions that
tend to the vertex. Because of the singularity of the reduced space
at the vertex, it seems to us that an investigation of motions
asymptotic to them is more naturally performed on the $\SO(3)$-reduced
system in $M_5$, and we leave it for a future work. 


The knowledge of the reduced dynamics in $M_4$ gives some information
on the properties of the motions of the unreduced system in $M_8$. In
particular, a rather complete description can be given for motions
that project over equilibria and periodic orbits of the reduced
system. Assume that a compact Lie group $G$ acts freely on a manifold
$\hat M$ and that $\hat X$ is a $G$-invariant vector field on $\hat
M$. Let $\pi:\hat M\to M:=\hat M/G$ be the quotient map and $X$ the
reduced vector field, which is $\pi$-related to $\hat X$. The
preimage under $\pi$ of an equilibrium of $X$ is called {\it relative
equilibrium} of $\hat X$ and  the preimage of a periodic orbit of $X$
is called {\it relative periodic orbit} of $\hat X$. The work of
\cite{field,krupa} proves that for each relative equilibrium (resp.
the relative periodic orbit) there exist an integer $0\le k\le\rank
G$ (resp. $1\le k\le1+\rank G$) and a vector $\o\in\reali^k$ such
that the relative equilibrium (resp. relative periodic orbit) is
fibered by $X$-invariant submanifolds diffeomorphic to $\toro^{k}$,
and the restriction of the flow of $\hat X$ to each of these
submanifolds is conjugate to the linear flow $\a\mapsto \a+t\o$
$\mathrm{mod}(2\pi)$ on $\toro^k$. We say that the flow in the
relative equilibrium or relative periodic orbit is {\it
quasi-periodic with $k$ frequencies}.

\begin{proposition}
\label{p:Rec} In $M_8$:
\ 
\bList
\item[i.] $\pi^{-1}(M_4^\mathrm{sing})$ is a union of relative
equilibria in each of which the flow of the unreduced system is
periodic (unless $p_4=0$ in which case the relative equilibrium
consists of equilibria).
\item[ii.] $\pi^{-1}(\mathcal{E}_4^\mathrm{reg})$ is a union of relative
equilibria in each of which the flow of the unreduced system is quasi-periodic with
$0\le k \le 2$ frequencies.
\item[iii.] In every relative periodic orbit, the flow of the
unreduced system is quasi-periodic with $1\le k\le 3$ frequencies.
\eList
\end{proposition}

\begin{proof} (i.) We have already remarked that in motions that project
onto the equilibria of the reduced system in the singular stratum
$M^\mathrm{sing}$ the ball stands on the vertex of the surface
$\tilde\Sigma$ and may have any vertical angular velocity.  (ii.) and
(iii.) follow from the fact that
the rank of $\SOTRE\times\SO(2)$ is 2.
\end{proof}


In view of Propositions \ref{p:RedDyn} and \ref{p:Rec},  in order to
reach a complete picture of the dynamics of the (reduced or
unreduced) system it is necessary to determine the reduced
equilibria in $\mathcal{E}_4^\mathrm{reg}$, and the motions asymptotic
to them, and the regions of the reduced space
$M_4^\mathrm{reg}\setminus\mathcal{E}_4^\mathrm{reg}$ in which the
(connected components of the) level sets of $(E,J)$ are compact and
those in which they are not. In the next section we make a first step
in this direction, looking for situations in which all the level sets
of $(E,J)$ are compact and hence the reduced dynamics in the 
complement of the set of the reduced equilibria and of their stable
and unstable sets is periodic, and the unreduced dynamics in the
complement of the set of relative equilibria and of their stable and unstable
sets is quasi-periodic.

\vskip4mm\noindent
{\it Remarks: } (i) The integrability by quadratures of the reduced
system was proved in \cite{BMK2002} by exploiting the existence of an
invariant measure and of the two Routh integrals and applying the Euler-Jacobi
theorem. However, this method cannot prove the periodicity of the
reduced dynamics. (At best, after replacing one of the Routh integrals
with the moving energy, it gives the weaker result that the reduced
dynamics is, after a time reparametrization, linear on tori of dimension two).

(ii) For the dynamics in relative equilibria and
relative periodic orbits in presence of a non compact symmetry group,
which also is of interest in nonholonomic mechanics, see
\cite{ashwin,FPZ2020}.

\subsection{Coercive profiles and quasi-periodicity of the unreduced
dynamics. }
\label{ss:superquadratic}

The simplest case in which {\it all} the level sets of $(E,J)|_{M_4}$ are
compact is when those of $E|_{M_4}$ are compact. Extending a result
in \cite{FGS2005} for the case $\O=0$ and for a convex profile, we
give some conditions that ensure this fact.

\begin{definition}
We say that the profile function $f$ is {\rm coercive} if 
$$
   \lim_{r\to +\infty} f(r) = +\infty
$$
and that it is {\rm asymptotically superquadratic} if 
\[
    \lim_{r\to +\infty} \frac{f(r)}{r^2} = +\infty \,.
\]
(Equivalently, $\lim_{p_1 \to +\infty} \psi(p_1)=+\infty$ in the first case
and $\lim_{p_1 \to +\infty} \frac{\psi(p_1 )}{p_1 } =
+\infty$ in the second).
\end{definition}

\begin{proposition}
\label{p:compactness}
The reduced moving energy $E|_{M_4}$ has all its level sets compact in
any one of the following two cases:
\bList
\item[(H1)] $\Omega=0$ and $f$ is coercive. 
\item[(H2)] $f$ is asymptotically superquadratic.
\eList
\end{proposition}

\begin{proof}
Since $E:\bR5\to\bR{}$ is continuous its level sets are closed and
we prove that their intersection with $M_4$ is bounded. 
Note that $\frac12p_2^2\psi'^2\ge0$ in all of $\bR5$ while, in
$M_4$,
$$
  p_1 \cF^2\psi'^2=\frac{p_1 \psi'^2}{1+2p_1 \psi'^2}\le\frac12
$$
and hence $-\Omega^2\mu p_1 \cF^2\psi'^2\ge
-\frac12 \mu \Omega^2$. Moreover, in $M_4$,
$p_2^2+p_3^2=4p_0 p_1 $ and hence $-|\Omega
p_3|\ge-2|\Omega|\sqrt{p_0 p_1 }$. Thus, in $M_4$,
$$
\begin{aligned}
  E
  & \ge \g\psi + p_0  + \frac12\mu p_4^2 +  \mu \Omega
  p_4 \cF
    -2\Omega  \sqrt{p_0 p_1 } + \Omega^2 \mu p_1
-\frac12\mu \Omega^2 
  \\
  & = \g\psi -\frac12\mu \Omega^2 -(1-\mu)\Omega^2p_1 
    + \mu \Big(\frac12p_4^2+ \Omega p_4\cF \Big) +
    \big(\sqrt{p_0 }-\Omega\sqrt{p_1 }\,\big)^2
  \\
  & = P + \mu Q + \big(\sqrt{p_0 }-\Omega\sqrt{p_1 }\,\big)^2
  \\   & \ge P + \mu Q
\end{aligned}
$$
where
$$
  P= \g\psi -\frac1{k+1}\Omega^2p_1  -\frac12\mu\Omega^2
  \,,\qquad 
  Q=\frac12p_4^2+ \Omega p_4\cF 
$$
(recall that $1-\mu=\frac1{1+k}$). In $M_4$, $0<\cF\le 1$ and
$Q\ge\frac12p_4^2-|\Omega p_4|\ge-\frac12\Omega=:Q_m$ is
bounded from below
and goes to $+\infty$ for $|p_4|\to+\infty$. Similarly, in
$M_4$, $p_1 \ge0$ and $P$ is bounded from below by a constant
$P_m\in\bR{}$. Moreover, if either
$\lim_{p_1\to+\infty}\psi(p_1)/p_1=+\infty$ (which happens
if $f$ is asymptotically superquadratic) 
or $\Omega=0$ and $\lim_{p_1\to+\infty}\psi(p_1)=+\infty$ 
(which happens if $f$ is coercive), then $P$ goes to
$+\infty$ for $p_1 \to+\infty$. 

Hence, in any level set $L_E$ of $E$, both $P$ and $Q$ are bounded
from below and from above. It easily follows from this that, in $L_E$, both
$p_1 $ and $p_4$ are bounded, so that $0\le p_1 \le c_1(E)$
and $|p_4|\le c_4(E)$ for some positive $c_1(E)$ and $c_4(E)$.
Since $(\sqrt{p_0 }-\Omega\sqrt{p_1 })^2\le E-P-\mu Q \le
E-P_m-\mu Q_m$, $p_0 $ is bounded as
well in $L_E$. Finally, from $p_2^2+p_3^2=4p_0     p_1$ it
follows that, in $L_E$, $p_2$ and $p_3$ are bounded as well.
\end{proof}

Since the map $J$ is continuous, under either of the two hypotheses of
Proposition \ref{p:compactness} the level sets of the map
$(E,J)|_{M_4}$ are compact and, as already pointed out, the reduced
dynamics is generically periodic and the unreduced dynamics is generically
quasi-periodic on tori of dimensions up to three.

\vskip4mm
{\it Remarks:} (i) For $\O=0$, Proposition \ref{p:compactness} was stated
in \cite{FGS2005} for convex profile functions, but a simple
inspection to the proof shows that what is there used is only the
coercivity of $f$, not its convexity.

(ii) When $\O\not=0$, the asymptotic superquadraticity of the profile
function is likely to be not only sufficient but also necessary for
the compactness of the level sets of $E|_{M_4}$. Indeed, for $p_2=p_4=0$
and large $p_1$, $E|_{M_4}$ is approximately equal to 
$
  \g\psi + \frac{p_3^2}{4p_1}-\O p_3+\mu\O^2p_1  
$
and hence, if $\psi$ goes to $+\infty$ not faster than $p_1$, to
$
  \frac{p_3^2}{4p_1}-\O p_3+\mu\O^2p_1  
$
whose level sets are hyperbolas (recall that $\mu<1$). The level sets
of the map $(E,J)|_{M_4}$ might nevertheless be compact. In fact, 
in Section \ref{s:paraboloid} we will show that this happens for the
parabolic profile $f(r)=b r^2$ with $b>0$; the same argument could be
easily applied to the case of the conic profile $f(r)=br$ with $b>0$.
A study of the compactness of the map $(E,J)|_{M_4}$ for a generic
profile is difficult because the functions $J_1$ and
$J_2$ are not explicitly known.

\section{Hamiltonization of the reduced system } 
\label{s:Ham}

\subsection{A rank-two Poisson structure. } 
The system formed by a sphere that rolls without sliding on a
surface of revolution which is at rest, namely our system for
$\Omega=0$ and a convex profile, has been one of the first---if not
even the very first---nonholonomic system with linear constraints and
a symmetry group for which it has been shown that the reduced system
is Hamiltonian with respect to a Poisson structure of rank two, with
the reduced energy as Hamiltonian  
\cite{BMK2002,ramos,FGS2005,BalseiroYapu}. 

We show here that the same remains true when $\Omega\not=0$, but with
the reduced moving energy, instead of the reduced energy, as
Hamiltonian. This is of interest for two reasons: From a geometrical
perspective, the very existence of Poisson structures for systems
with {\it affine} (rather than linear) constraints was so far unknown,
except in the very special case of the Veselova system \cite{GN2007}.
And from a
dynamical perspective, it helps enlightening some aspects of the
dynamics of the reduced system, which turns that of a (family of)
Hamiltonian systems with one degree of freedom which are of
mechanical type (hence, also Lagrangian).

We limit ourselves to consider the reduced system in the subset of
the regular stratum $M_4^\mathrm{reg}$ where $p_1\not=0$. As we have
already noticed, $M_4^\mathrm{reg} \setminus\{p_1=0\}$ is diffeomorphic to
$$
  M_4^\circ := \reali_+\times\reali^3 \ni (p_1,p_2,p_3,p_4) \,,
$$
with diffeomorphism $M_4^\circ \to M_4^\mathrm{reg}\setminus \{p_1=0\}$ 
given by
$
  (p_1,p_2,p_3,p_4)
  \mapsto 
  \Big(\frac{p_2^2+p_3^2}{4p_1},p_1,p_2,p_3,p_4\Big)
$.
We thus pull back the entire description to $M_4^\circ$, and we
denote with a superscript~${}^\circ$ the pull-backed objects on
$M_4^\circ$. In this way, the restriction to $M_4^\mathrm{reg}\setminus
\{p_1=0\}$ of the vector field $X=(X_0,X_1,\dots,X_4)$ in $\bR5$
given by \eqref{X} becomes the vector field $X^\circ$ in $M_4^\circ$
with components 
\begin{equation}\label{EqRid-2}
   X^\circ_i = X_i\big|_{p_0=\frac{p_2^2+p_3^2}{4p_1}}
   \,,\qquad i=1,\ldots,4 \,. 
\end{equation}
Similarly, the reduced moving energy \eqref{E} becomes the function
$E^\circ:M_4^\circ\to\reali$ given by
$$
  E^\circ \;=\; \frac{p_2^2}{4p_1\cF^2} + \g\psi + \frac{p_3^2}{4p_1}
    + \frac12\mu p_4^2
    \;+\;
  \Omega\big(\mu p_4 \cF -p_3\big)
  \;+\; \Omega^2 \mu p_1  \big(1-\cF^2 \psi'^2 \big) \,.
$$
The representative $J^\circ:M_4^\circ\to\rdue$ of $J|_{M_4^\mathrm{reg}
\setminus\{p_1=0\}}$ has the same expression \eqref{J} as $J$, but we
prefer using the symbol $J^\circ$ to stress that we are working in a
subset of $M_4^\mathrm{reg}$, and with a different parametrization.

\begin{proposition}
\label{p:Poisson}
Consider the bivector
$$
 \Lambda \;:=\; 
    2p_1\cF^2  \partial_{p_2}\wedge \Big(
    \partial_{p_1} 
    + (G_3p_4+\Omega g_3) \partial_{p_3} 
    + (G_4p_3+\Omega g_4) \partial_{p_4} 
    \Big) 
$$
on $M_4^\circ$. Then:
\bList
\item[i.] $X^\circ =\Lambda(dE^\circ,\cdot )$.
\item[ii.] $\Lambda$ is a rank-two Poisson tensor on $M_4^\circ$.
\item[iii.] The two components of $J^\circ$ are Casimirs of
$\Lambda$.
\eList
\end{proposition}

\begin{proof} (i.) In the dense subset of $M_4^\circ$ where $p_2\not=0$,
$\Lambda = \frac{2p_1}{p_2} \cF^2 \partial_{p_2}\wedge X^\circ$. Since
$L_{X^\circ}E^\circ=0$, in such a subset $\Lambda(dE^\circ,\cdot) =
(\frac{2p_1}{p_2}\cF^2\partial_{p_2}E^\circ)X^\circ  =
\frac{2p_1}{p_2}\cF^2\big( \frac{p_2}{2p_1} + p_2\psi'^2\big)X^\circ
= X^\circ$. By continuity, this is true in all of $M_4^\circ$.

(ii.) The characteristic distribution of the bivector
$\Lambda$ is spanned by the two vector fields $\partial_{p_2}$ and
$\partial_{p_1} + (G_3p_4+\Omega g_3) \partial_{p_3}
+ (G_4p_3+\Omega g_4) \partial_{p_4}$, which are everywhere linearly
independent. Thus $\Lambda$ has everywhere rank two and the associated
Poisson brackets trivially satisfy the Jacobi identity, so that
it is Poisson. 

(iii.) From \eqref{J}, $J^\circ=U^{-1}(\hat p+\Omega g)$ with $\hat
p=\Big(\begin{matrix}p_3\\ p_4\end{matrix}\Big)$.
Recalling that $\partial_{p_1}U^{-1} = -U^{-1}G$ we have, for each $i=1,2$,
$$
   \partial_{p_1}J_i^\circ = -[U^{-1}G\hat p+\Omega U^{-1}g]_i =
   -[U^{-1}Gp_3e_1+U^{-1}Gp_4e_2 + \Omega U^{-1}g_3e_1
    + \Omega U^{-1}g_4e_2]_i 
$$
where $e_1=\Big(\begin{matrix}1\\0\end{matrix}\Big)$ and
$e_2=\Big(\begin{matrix}0\\1\end{matrix}\Big)$. Moreover,
$$
  (G_3p_4+\Omega g_3)\partial_{p_3}J^\circ_i = 
  (G_3p_4+\Omega g_3)[U^{-1}e_1]_i = 
  [U^{-1}p_4 G_3 e_1+\Omega U^{-1}g_3 e_1]_i = 
  [U^{-1}Gp_4 e_2+\Omega U^{-1}g_3 e_1]_i 
$$
and, similarly, $(G_4p_3+\Omega g_4)\partial_{p_4}J^\circ_i =  [U^{-1}Gp_3
e_1+\Omega U^{-1}g_4 e_2]_i$. Hence $\Lambda(dJ^\circ_i,\cdot)=0$. 
\end{proof}

We point out that, for $\O\not=0$, the origin of the rank-two
Poisson structure $\Lambda$ is not clear. There are two
possible approaches:

1. There exists an almost-Poisson formulation of nonholonomic
mechanical systems with {\it linear} constraints and
Lagrangian without gyrostatic terms  \cite{BS, VanDerSchaft}. In
presence of symmetry---and under suitable hypotheses---this
almost-Poisson structure induces a Poisson structure on the reduced
space, that makes the reduced system Hamiltonian with the energy as
Hamiltonian \cite{BN2011, Balseiro, GM2018, balseiro2017,BalseiroYapu}.
A similar theory for the case of {\it
affine} constraints (or, equivalently, for Lagrangians with
gyrostatic terms) does not exist yet. We speculate that such an
extension might exist, particularly if the reduced moving energy is
`kinematically interpretable' in the sense of \cite{FNS2018}.  

2. In \cite{FGS2005}, it is shown that every dynamical system with
periodic flow possesses (infinitely many) rank-2 Poisson formulations,
suggesting a dynamical origin of these structures. This point of view
may account for the existence of $\Lambda$ in the case of coercive
profiles, but not in general. It is possible that the approach of
\cite{FGS2005} could be
extended by using the existence of three first integrals, even if
their level sets are not compact.

\subsection{The $J^\circ$-restricted reduced systems. }

The symplectic leaves of the Poisson manifold $(M_4^\circ,\Lambda)$
are the level sets of the Casimir map $J^\circ:M_4^\circ\to\rdue$.
Clearly, this map is surjective and, for any $j\in\rdue$, the level
set $M_2^j:=(J^\circ)^{-1}(j)$ is given by
$$
  M_2^j
  \,=\,
  \big\{(p_1,p_2,p_3,p_4) \in M_4^\circ \,:\, 
        p_3= \tilde p_3(p_1,j),\, p_4=\tilde p_4(p_1,j) \big\} \,,
$$ 
with $\tilde p_3$ and $\tilde p_4$ defined by \eqref{tildep3p4}, 
and is a submanifold of $M_4^\circ$ diffeomorphic to $\reali_+\times
\reali\ni(p_1,p_2)$. The Poisson structure $\Lambda$ induces a
symplectic form $\o_j$ on each symplectic leaf $M_2^j$, and the
restriction of $X^\circ$ to $M_2^j$ equals the vector field
$\o_j^\flat\big(dE^\circ|_{M_2^j}\big)$, namely, the 
$\o_j$-Hamiltonian vector field whose Hamiltonian is the restriction 
of the reduced moving energy $E^\circ$ to $M_2^j$. 

If we use $(p_1,p_2)$ as coordinates on $M_2^j$, then
$$
  \o_j(p_1,p_2) = \frac1{2p_1\cF^2}\, dp_2\wedge dp_1 
$$
and $E^\circ|_{M_2^j}(p_1,p_2)= \frac12\,\frac{p_2^2}{2p_1\cF(p_1)^2}
+ W_{\Oj}(p_1)$ with ``effective potential''
$$
  W_{\Oj} = \g\psi + \frac{\tilde p_{3,j}^2}{4p_1}
   + \frac12\mu\tilde p_{4,j}^2 
   +\Omega\big(\mu \tilde p_{4,j}\cF - \tilde p_{3,j}\big)
   +\Omega^2 \mu p_1(1-\cF^2\psi'^2) \,.
$$
where $\tilde p_{3,j}$ and $\tilde p_{4,j}$ stand for $\tilde
p_3(\cdot,j)$ and $\tilde p_4(\cdot ,j)$.  If we pass to the
(Darboux) coordinates $(Q,P)=\big(p_1,\frac{p_2}{2p_1\cF^2}\big) \in
\reali_+\times\reali$ on $M_2^j$, then the symplectic 2-form $\o_j$
becomes $dP\wedge dQ$ and $E^\circ|_{M_2^j}$ becomes $\frac12
{2p_1\cF^2} {p_2^2}+ W_{\Oj}(p_1)$. Thus, the restriction of the
reduced system to each symplectic leaf can be regarded as a
Hamiltonian system that describes a one-degree-of-freedom mechanical
(holonomic) system on the cotangent bundle $T^*\reali_+\ni(Q,P)$ of
the configuration space $\reali_+\ni Q=p_1=r^2/2$. Equivalently, this
can be regarded as a Lagrangian system on $T\reali_+\ni(Q,\dot
Q)=(p_1,\dot p_1)$ with `natural' Lagrangian $\frac12 \frac{\dot
p_1^2}{2p_1\cF^2}-W_{\Oj}(p_1)$. To allow for easier interpretation,
we prefer switching to the coordinates $(r,\dot r)$. Correspondingly,
we reverse to the original profile function $f(r)$ and we use the two
functions
$$
  \overline p_{i,j}(r):=\tilde p_{i,j}\Big(\frac{r^2}2\Big)
  \,,\qquad i=3,4 \,.
$$

\begin{proposition}
\label{p:Lagr}
The restriction of the reduced equations \eqref{EqRid} to any level set
$M_2^j$ of the two reduced Routh integrals, written in
coordinates $(r,\dot r)\in T\reali_+$, is the Lagrangian system with
Lagrangian
\beq{Lagr}
  \frac 12 F(r)^2\dot r^2 - V_{\Oj}(r)
\eeq
with the effective potential
\beq{Vj}
  V_{\Oj} = \g f
   + \frac{\overline p_{3,j}^2}{2r^2} + \frac12\mu\overline p_{4,j}^2
   +\Omega\Big(\mu\frac{\overline p_{4,j}}{F}-\overline p_{3,j}\Big)
   +\frac12 \mu \Omega^2 \Big(r^2 -\frac{f'^2}{F^2}\Big) \,.
\eeq
\end{proposition}

\section{Reduced equilibria in $\mathcal{E}_4^\mathrm{reg}$. } 
\label{s:RedEq}

\subsection{The reduced equilibria in $\mathcal{E}_4^\mathrm{reg}$. } 
\label{ss:RedEq-1}

In this section we study the reduced equilibria in $\mathcal{E}_4^\mathrm{reg}$. Since at an
equilibrium with $p_1=0$ (namely $r=0$) it is necessarily
$p_0=p_2=p_3=0$, all equilibria with $p_1=0$ belong to
$M_4^\mathrm{sing}$. Therefore, $\mathcal{E}_4^\mathrm{reg}\subset
M_4^\mathrm{reg}\setminus\{p_1=0\}$ and
for
easier interpretation we may work in $M^\circ_4$ with the coordinates
$(r,v_r,v_\theta,\o_n)$ (which in the Appendix is called $\widehat
M_4^\circ$; recall that $p_1=\frac{r^2}2$, $p_2=rv_r$,
$p_3=p_1v_\theta$, $p_4=\o_n$). Obviously, $v_r=0$ at all reduced
equilibria.

\begin{proposition}
\label{p:RE-1}
For any $\O\in\reali$ and $\bar r>0$, the reduced
equilibria with $r=\bar r$ form three disjoint families:
\bList
\item[(RE1)] If $f'(\bar r)=0$, the 1-parameter family
$\cP_1(\bar r, \o_n,\O) = (\bar r,0,\O\mu,\o_n)$, 
$\o_n\in\reali$.
\item[(RE2)] If $f'(\bar r)=0$ and $\O\not=0$, also the 1-parameter
family $\cP_2(\bar r, \o_n,\O) = (\bar r,0,0,\o_n)$, 
$\o_n\in\reali$.
\item[(RE3)] If $f'(\bar r)\not=0$, the 1-parameter family 
$\cP_3(\bar r,v_\theta,\O) = \big(\bar r,0,v_\theta,
\tilde\o_n(\bar r,v_\theta,\O)\big)$, $v_\theta\not=0$, where
\beq{omegatilden}
 \tilde\o_n(r,v_\theta,\O)
 :=
 \frac{r v_\theta}{\mu f'(r)}
 -
 \frac{\g}{\mu v_\theta}
 -
 \O\Big(\frac r{f'(r)}+\frac1{F(r)}\Big) \,.
\eeq
\eList
\end{proposition}

\begin{proof} The equilibria of the reduced vector field $X^\circ$
in $M_4^\circ$ are the points $(p_1,p_2,p_3,p_4)$ where $p_2=0$ and
$X_2\big(\frac{p_3^2}{4p_1},p_1,0,p_3,p_4\big)=0$. Since $\cF$ never
vanishes, the latter condition is
\beq{RE}
  \frac{p_3^2}{2p_1} -\mu p_3p_4\psi'(p_1) -2\g p_1\psi'(p_1) 
  -\O\mu p_3\big(1+\psi'(p_1)\cF(p_1)\big)
  = 0 \,.
\eeq
If $\psi'(p_1)=0$ this condition becomes
$$
  \frac{p_3^2}{2p_1} -\O\mu p_3 = 0 
$$
and has the solutions $p_3=2\O\mu p_1$ and $p_3=0$, which give the
reduced equilibria of types RE1 and RE2, respectively.  If
$\psi'(p_1)\not=0$, then \eqref{RE} does not have any solution with
$p_3=0$. Equation \eqref{RE} can then be solved for $p_4$, obtaining
\begin{equation}
\label{RE-2}
  p_4 = -\frac{2\g}\mu\,\frac{p_1}{p_3} + \frac{p_3}{2\mu p_1\psi'(p_1)}
        -\frac\O{\psi'(p_1)}-\O\cF(p_1) \,.
\eeq
In the coordinates $(r,v_r,\dot\theta,\o_n)$ this is
$\o_n=\tilde\o_n(r,v_\theta,\O)$. \end{proof}

\noindent
Thus, for each $r>0$ and $\O\in\reali$ there are one or two
1-parameter families of reduced equilibria with those $r$ and $\O$. 
Families RE1 and RE2 are parametrized by $\o_n\in\reali$, while
family RE3 is parametrized by $v_\theta\not=0$. For fixed $r$ and
$\O$, the curve $\o_n=\tilde\o_n(r,v_\theta,\O)$ in the plane
$(v_\theta,\o_n)$ has two branches, one in the half plane
$v_\theta>0$ and one in the half plane $v_\theta<0$. When $\O=0$
these two branches are symmetrical with respect to the origin. The
qualitative properties of these curves depend on the sign of $f'(r)$,
and are shown in Figure 2 for $\O=0$; a nonzero $\O$ shifts both
branches up or down, depending on the signs of $\O$ and of the
quantity $\frac r{f'}+ \frac 1F$ and has no effect on them if the latter
quantity vanishes.
Curiously, if $f'(r)>0$ then there are exactly two reduced 
equilibria with $\o_n=0$. 


\begin{figure}[h]
\begin{center}
{\small
{\scalebox{1.}{\includegraphics*{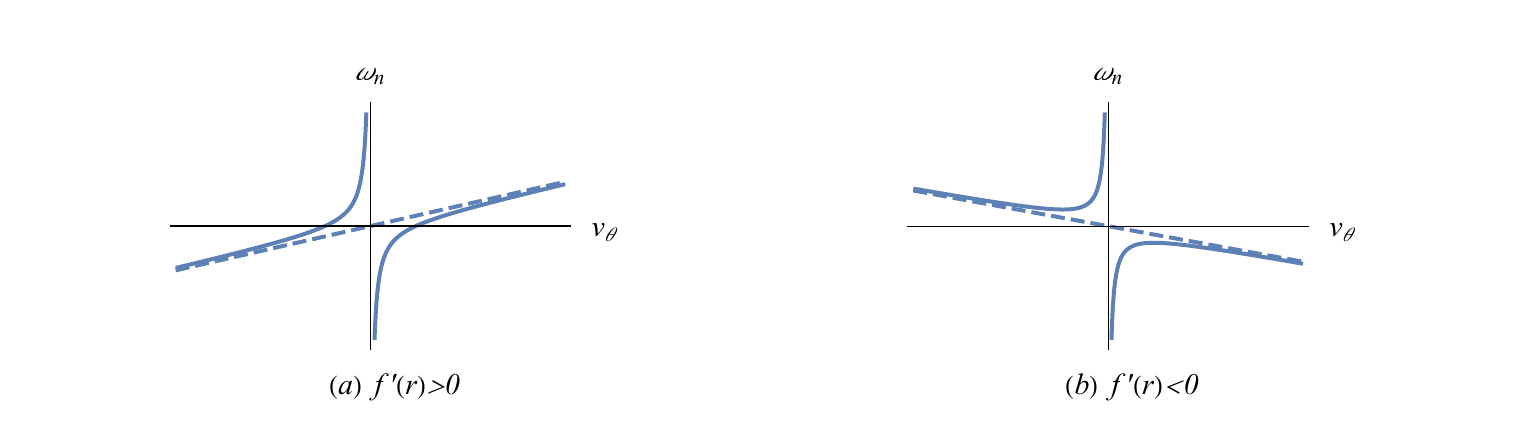}}}
}
\caption{\small The two branches of the RE3 equilibria
$\cP_3(r,v_\theta,0)$ in the plane $(v_\theta,\o_n)$ for $\O=0$.
The dotted line is the asymptote $\o_n=\frac r{\mu f'(r)}v_\theta$.}
\end{center}
\end{figure}


A more difficult question is which reduced equilibria are present for
any given value of $J^\circ=(J^\circ_1, J^\circ_2)$. This depends in a
non obvious way on the profile of the surface $\Sigma$ and on $\O$,
given that the map $J^\circ$ depends on them, and can be investigated,
numerically if not analytically, on a case by case basis. The case of
an upward half-cone  was studied in \cite{BIKM2019}. The case of an
upward paraboloid is studied in Section \ref{s:paraboloid}. 

\vskip4mm
{\it Remarks: } 
(i) The reason why, when $\O=0$, we consider the reduced equilibria
$(\bar r,0,0,\o_n)$ as part of the family RE1, instead of RE2, is
because of their stability properties. 

It follows from \eqref{ReflSimm} that, if $(r,0,v_\theta,\o_n)$ is
an equilibrium of the reduced system for a certain value of
$\O$, then $(r,0,-v_\theta,-\o_n)$ is an equilibrium of the reduced
system for $-\O$, and they have the same stability properties. (This
can also be checked with \eqref{RE} and with the formulas of
Proposition \ref{p:RE-3}). We may therefore restrict our study of the
reduced equilibria to the case $\O\ge0$.

When $\O=0$, the invariance of $X$ under the reflection $C$ as in
\eqref{C} implies that if
$(r,0,v_\theta,\o_n)$ is a reduced equilibrium then so is
$(r,0,-v_\theta,-\o_n)$ and they have the same stability properties. Note
that, by \eqref{ReflRouth}, if one of them belongs to $M_2^j$,
then the other belongs to $M_2^{-j}$. When $\O=0$ we may thus restrict
ourselves to study reduced equilibria for $j_1\in\reali$, $j_2\ge0$.

\subsection{Motions in relative equilibria. }
\label{ss:RelEq}

Motions in all relative equilibria in $M_8$
consist of a uniform rotation of the center of mass of the ball on a
parallel (hence, a horizontal circle) of the surface $\Sigma$, and of
a uniform rotation of the ball around the axis normal to $\Sigma$
(which changes periodically with the same frequency as the center of
mass). See also Proposition \ref{p:Rec}.

By Proposition \ref{p:RE-1}, there are three families of relative
equilibria, which we call with the same names of the reduced
equilibria onto which they project, and there is at least one such
family on any parallel of $\Sigma$. 
For each $\O\in\reali$:
\bList
\bull Relative equilibria of type RE1 consist of
motions in which the center of mass of the ball uniformly moves (if
$\O\not=0$) or stands (if $\O=0$) on a horizontal `critical' parallel
of the surface $\Sigma$. At these points the normal vector $n$ is
vertical. Note that, since $\frac12<\mu<1$, the angular velocity
$v_\theta=\O\mu$ of the center of mass is smaller than that of the
surface. Thus, the ball either rolls (if $\O\not=0$) 
or stands (if $\O=0$) on the corresponding critical parallel of the surface
$\tilde\Sigma$, and at the same time rotates around its vertical axis
with any constant angular velocity $\o_\z=\o_n$. 

\bull In relative equilibria RE2, $v_\theta=0$ and the center of
mass of the ball stands still in space. Correspondingly, the ball
rolls uniformly on a critical parallel of the surface $\tilde\Sigma$.
Here too, the ball may rotate with any constant angular velocity
$\o_n=\o_\z$ around its vertical axis.

\bull
In relative equilibria of type RE3 the ball rolls along a
non-critical parallel of the surface $\tilde\Sigma$, with any nonzero
$v_\theta$.

\eList

\vskip4mm
{\it Example. } The case of a ball on a plane ($\psi=0$) is well
known and elementary \cite{earnshaw,NF}. The equations of motion for
the $\SO(3)$-reduced system in $M_5\ni(x,\dot x,\o_\z)$ are $\ddot x
= -\mu\O \dot y$, $\ddot y = \mu\O \dot x$, $\dot \o_z = 0$
(Equations (5.44) in \cite{NF}). $\o_z=\o_n$ is constant. If $\O=0$ the center of mass
moves on a straight line or stands still. For $\O=0$ the solution
with initial conditions $(x_0,y_0,\dot x_0,\dot y_0)$ is
$$
  x(t) =
  x_0 - \frac{\dot y_0}{\mu\O}
  + \frac{\dot y_0}{\mu\O} \cos(\mu\O t)
  + \frac{\dot x_0}{\mu\O} \sin(\mu\O t)
  \,,\quad
  y(t) =
  y_0 + \frac{\dot x_0}{\mu\O}
  + \frac{\dot y_0}{\mu\O}\sin(\mu\O t)
  - \frac{\dot x_0}{\mu\O} \cos(\mu\O t)
$$
and the center of mass moves along a circle. According to Proposition
\ref{p:RE-1} the $S^1$-reduction to $M_4$ of this system in $M_5$ has two
families of reduced equilibria at any distance $r$ from the origin,
one of type RE1 and one of type RE2. The lift to $M_5$ of the reduced
equilibria of type RE1 are motions with $\dot
x_0=-\mu\O y_0$, $\dot y_0=\mu\O x_0$ with nonzero $(x_0,y_0)$: the ball
spins with any $\o_z$ around its center of mass, that moves along a
circle centered at the origin. The lift to $M_5$ of the reduced equilibria of
type RE2 are motions with initial conditions $\dot x_0=\dot y_0=0$:
the ball spins with any $\o_z$ around its center of mass, that stands
still in space. 

\vskip4mm
{\it Remarks: } (i) Relative equilibria of type RE2 resemble certain
motions of a ball on a rotating umbrella produced in the Japanese
`turning umbrella' (kasamawashi) art. In some of these performances,
an umbrella is kept in uniform rotation about its inclined axis, and
a ball rolls on its surface in such a way to remain fixed in space.
At each instant, the ball touches a point of the umbrella whose
tangent plane is horizontal. The difference with our treatment is
that, due to the inclination of the umbrella, that system is not
invariant under rotation about the vertical. We will come back on
this system in a future work.

(ii) In view of the example of the ball on the rotating plane, the
existence of the reduced equilibria of types RE1 and RE2 can be
regarded as obvious.  However, the {\it stability} of these
equilibria depends on the surface profile, see next Section.

\section{(Leafwise) stability of the reduced equilibria}

\subsection{Leafwise-stability. }
\label{ss:RE}

We study now the stability of the reduced equilibria---where 
`stability' is relative to the restriction of the reduced
system to a level set $M_2^j$ of the map $J^\circ$. In order to avoid
ambiguities on this point, we introduce the following terminology:

We say that an equilibrium of the reduced system is {\it
leafwise-stable} ({\it leafwise-unstable})  if it is a
Lyapunov-stable (Lyapunov unstable) equilibrium of the restriction of
the reduced system to the level set $M_2^j$ of the map $J^\circ$ to
which it belongs. (`Leafwise' refers, of course, to the symplectic
leaves of the Poisson structure of $M_4^\mathrm{reg}$).

Leafwise-stability of a reduced equilibrium does not imply its
stability as equilibrium of the reduced system in $M_4^\mathrm{reg}$,
because motions nearby might run away with small but nonzero
$v_\theta$. However, it implies the $\SOTRE\times\SO(2)$-orbital
stability of the motion in the corresponding relative equilibria of
the unreduced system.

By Proposition \ref{p:Lagr}, a reduced equilibrium in $M_2^j$ is a 
point $(r,\dot r=0)\in M_2^j$ with $r$ a critical point of
$V_{\Oj}$ and, given the Lagrangian nature of the restriction of the
reduced system to $M^j_2$, it is leafwise-stable if $V''_j(r)>0$,
leafwise-unstable if $V''_j(r)<0$. This leads to the following
conditions:

\begin{proposition} 
\label{p:RE-3}
For any $r>0$ and $\O\in\reali$:
\bList
\item[i.] A reduced equilibrium $\cP_1(r,\o_n,\O)$ of type RE1 is
leafwise-stable if $S_1(r,\O)>0$ and leafwise-unstable if $S_1(r,\O)<0$,
where
\beq{RE1-stab}
   S_1(r,\O):=\mu^2\O^2+\g f''(r) \,.
\eeq

\item[ii.] A reduced equilibrium $\cP_2(r,\o_n,\O)$ of
type RE2 (with $\O\not=0$) is leafwise-stable if $S_2(r,\o_n,\O)>0$ 
and leafwise-unstable if $S_2(r,\o_n,\O)<0$, where
\beq{RE2-stab}
  S_2(r,\o_n,\O) := \mu^2\O^2+\big(\g+\mu^2\o_n\O+\mu^2\O^2\big)f''(r) 
  \,.
\eeq

\item[iii.] A reduced equilibrium $\cP_3(r,v_\theta,\O)$ of type RE3
is leafwise-stable if $S_3(r,v_\theta,\O)>0$ and leafwise-unstable if 
$S_3(r,v_\theta,\O)<0$ where
\beq{RE3-stab}
  S_3(r,v_\theta,\O) := \D_0(r,v_\theta) + \O \D_1(r,v_\theta) 
\eeq
with
$$
  \D_0(r,v_\theta)
  = 
  \D_{00}(r) + \D_{02}(r)v_\theta^2 + \D_{04}(r)v_\theta^4
  \,,\qquad
  \D_{1}(r,v_\theta) 
  = 
  \D_{11}(r)v_\theta 
$$
and 
\begin{eqnarray*}
  \D_{00}
  \ugarr
  \g^2 f'f''  
  \,,\qquad
  \D_{02} = 2\g F^2 f' 
  \,,\qquad
  \D_{04} = (1+\mu f'^2)rF^2 + (1-\mu)r^2f'f'' \,,
  \\
  \D_{11}
  \ugarr
  \g\mu\big(rf'' - F^2 f'\big) \,.
\end{eqnarray*}
\eList
\end{proposition}

\begin{proof} 
Let $p_1=r^2/2$. The equilibrium belongs to a level set $M_2^j$ of
$J$ and, as remarked, it is leafwise-stable if $W''_\Oj(p_1)>0$ and
leafwise-unstable if $W''_j(p_1)<0$. Computing $W''_\Oj(p_1)$ using
$\tilde p_{3,j}'=G_3\tilde p_{4,j}+\O g_3$ and  $\tilde p_{4,j}'=G_4\tilde
p_{3,j}+\O g_4$ we obtain $W''_j=D_0+\O D_1+\O^2 D_2$ with
\begin{eqnarray*}
  D_0
  \ugarr
  \g\psi''
  +
  \mu^2p_4^2\Big(\frac{\psi'}{2p_1}+\psi''\Big) \cF^2 \psi' 
  +
  \mu\frac{p_3p_4}{2p_1^2} 
  \big(p_1 (1-2\cF^2)\psi''-(1+\cF^2)\psi'\big)
  \\
  \plusarr
  \frac{p_3^2}{2p_1^3}
  \big( 1 + 2p_1\psi'^2 + \mu p_1^2(\psi'^3-\psi'')\psi'\big)  \cF^2
  \,,
  \\
  D_1
  \ugarr
  \mu^2\frac{p_4}{p_1}
  \Big( \big(1+\cF\psi'+p_1\psi'^2\big)\psi'
       + p_1\big(1+2\cF\psi'\big)\psi''
  \Big) \cF^2
  \\
  \ \quad\plusarr
  \mu\frac{p_3}{2p_1}
  \Big(
     \big(1 + 2p_1\cF\psi' -2\cF^2 \big)\cF\psi''
     - \frac1{p_1}(1+\cF^2)(1+\cF\psi') - \cF^2 \psi'^2 
   \Big)\,,
   \\
   D_2
   \ugarr
   \frac{\mu^2}{2p_1}
   \big(1+\cF\psi'\big) 
   \big(1+\cF^3\psi'+2p_1\cF^3\psi''\big) \,.
\end{eqnarray*}
(Here and below in this proof $p_3$ and $p_4$ stand, respectively,
for $\tilde p_{3,j}$ and  $\tilde p_{4,j}$).

(i.) If $\psi'(p_1)=0$ then $\cF(p_1)=1$ and, if moreover $p_3=2\O\mu p_1$,
then $W''_\Oj(p_1)= \g \psi''(p_1) + \frac{\mu^2\O^2}{2p_1}$. If
$\psi'(p_1)=0$ then $\psi''(p_1)=\frac{f''(r)}{2p_1}$ and reversing to the
coordinate $r$ this gives the stated result.

(ii.) If $\psi'(p_1)=0$ and $p_3=0$ then $W''_\Oj(p_1)=
\frac1{2p_1}\mu^2\O^2 + \big(\g + \mu^2
p_4\O+\mu^2\O^2\big)\psi''(p_1)$. 

(iii.) At the reduced equilibria of type RE3, $p_3\not=0$ and $p_4$ is
given by \eqref{RE-2}. Inserting this expression in the formulas above
gives $W''_j(p_1)=d_0+\O d_1$ with
\begin{eqnarray*}
  d_0
  \ugarr
  2\g^2\frac{p_1}{p_3^2}\big(\psi'+2p_1\psi'')\psi'\cF^2 
  +
  \g \frac{\psi'}{p_1}
  +
  \frac{p_3^2}{2p_1}\Big(\frac1{4p_1^2}+\frac{\psi'^2+\mu p_1\psi'^4}{p_1} 
            + (1-\mu)\psi'\psi''\Big) \cF^2
  \\
  d_1
  \ugarr 
  2\g\mu\frac{p_1}{p_3} \big(\psi''-\psi'^3)\cF^2 \,.
\end{eqnarray*}
Up to the change of coordinates, $\D_0=F^2 r^3v_\theta^2d_0$ and 
$\D_1 = F^2 r^3v_\theta^2 d_1$.
\end{proof}

We now draw some consequences from Proposition \ref{p:RE-3}. Of
special interest is the effect of the rotation of the surface on the
properties of leafwise-stability of the reduced equilibria. However, also
the case $\O=0$ is of interest because it has been so far
investigated only very partially \cite{routh,hermans,zenkov}.
As remarked, we may restrict the analysis to the case $\O\ge0$.

\subsection{Leafwise-stability of RE1 reduced equilibria. }
\label{ss:stab-RE1}

The properties of leafwise-stability of the reduced equilibria of type
RE1 are read without any difficulty from the expression of the function
$S_1$ as in \eqref{RE1-stab}. Assume $f'(r)=0$. 

As it might be expected, when $\O=0$ all reduced equilibria
$\cP_1(r,\o_n,0)$, $\o_n\in\reali$, are leafwise-stable if $f''(r)>0$
and leafwise-unstable if $f''(r)<0$. For a given $\O\not=0$,
$\cP_1(r,\o_n,\O)$ is leafwise-stable if
$f''(r)>-\frac{\mu^2}\g\O^2$. 

Since \eqref{curvatura} implies $f''(r)>-1$ at any critical point $r$ of
$f$, for
$|\O|>\frac{\sqrt\g}\mu$ all reduced equilibria of type RE1 are
leafwise-stable. The rotation of the surface has thus a stabilizing
effect on reduced equilibria of type RE1 (a sort of  `gyrostatic
stabilization').

Note that the properties of leafwise-stability of reduced equilibria
of type RE1 are independent of the angular velocity $\o_n=\o_z$ of the
ball.

\subsection{Leafwise-stability of RE2 reduced equilibria. }
\label{ss:stab-RE2}

According to the choice we made, reduced equilibria of type RE2 are
defined only for $\O\not=0$, and we consider them only for $\O>0$.
Proposition \ref{p:RE-3} implies that

\begin{proposition}
\label{p:stab-RE2}
Assume $f'(r)=0$. Then, for any $\O>0$: 
\bList
\item[i.] If $f''(r)=0$, all reduced equilibria $\cP_2(r,\o_n,\O)$,
$\o_n\in\reali$, are leafwise-stable.
\item[ii.] If $f''(r)>0$, $\cP_2(r,\o_n,\O)$ is
leafwise-stable if
$$
   \o_n > -\frac{1+f''(r)}{f''(r)}\O- \frac{\g}{\mu^2\O}
$$ 
and leafwise-unstable if $\o_n$ satisfies the opposite inequality.
\item[iii.]  If $f''(r)<0$ (hence $|f''(r)|<1$), $\cP_2(r,\o_n,\O)$
is leafwise-stable if 
$$
   \o_n < \frac{1-|f''(r)|}{|f''(r)|}\O- \frac{\g}{\mu^2\O}
$$ 
and leafwise-unstable if $\o_n$ satisfies the opposite inequality.
\eList
\end{proposition}

Thus, the rotation of the surface has a stabilizing effect also on
the reduced equilibria of type RE2: they all become leafwise-stable
for $\O\to+\infty$.  

The regions of leafwise-stability and leafwise-instability of these
reduced equilibria in the half-plane $(\O,\o_n)\in\reali_+\times
\reali$ are depicted in Figure 3 for the cases in which
$f''(r)\not=0$. Note that, in these cases, the stability properties
depend also on the angular velocity $\o_n=\o_\z$ with which the ball
rotates about its vertical axis.

\begin{figure}[h]
\begin{center}
{\small
{\scalebox{1.}{\includegraphics*{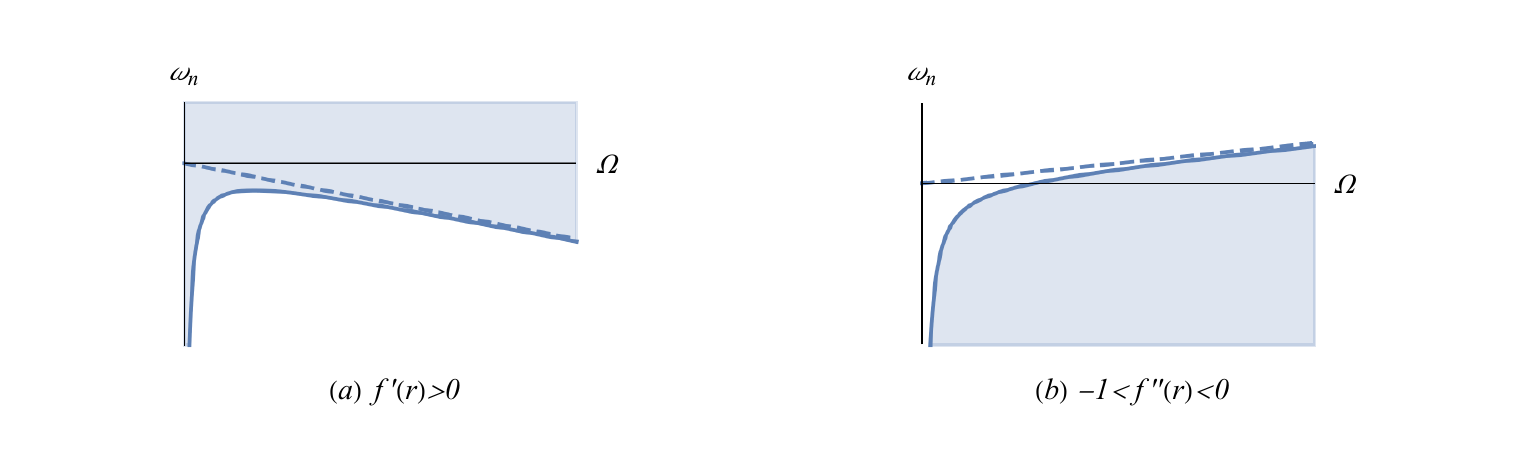}}}
}
\caption{\small Bifurcation diagrams for the reduced equilibria of type RE2.
Reduced equilibria are leafwise-stable in the shaded regions and
leafwise-unstable in the unshaded regions. The boundary of the two
regions is the curve $\o_n=-\frac{1+f''(r)}{f''(r)}\O-
\frac{\g}{\mu^2\O}$. The dashed curve is the asymptote
$\o_n=-\frac{1+f''(r)}{f''(r)}\O$.}
\end{center}
\end{figure}


\subsection{Leafwise-stability of RE3 reduced equilibria. }
\label{ss:stab-RE3}

Reduced equilibria of type RE3 exhibit more complex bifurcation
scenarios than those of types RE1 and RE2. As above, we may assume $\O\ge0$. 

First we note that, for large $\O$, the surface rotation may have
either a stabilizing or a de-stabilizing effect on these reduced
equilibria, depending on the direction in which the ball
moves along the surface's parallel, or even (in non-generic but
nontrivial cases) no effect at all:

\begin{proposition} 
\label{p:RE-6}
Consider $r>0$ such that $f'(r)\not=0$ and $v_\theta\not=0$.
\bList
\item[i.] If $\D_{11}(r)=0$, then the properties of
leafwise-stability of $\cP_3(r,v_\theta,\O)$ are independent of~$\O$. 
\item[ii.] If $\D_{11}(r)>0$, then for $\O$ large enough
$\cP_3(r,v_\theta,\O)$ is leafwise-stable if
$v_\theta>0$ and leafwise-unstable if $v_\theta<0$. 
\item[iii.] If $\D_{11}(r)<0$, then for $\O$ large enough
$\cP_3(r,v_\theta,\O)$ is leafwise-stable if
$v_\theta<0$ and leafwise-unstable if $v_\theta>0$. 
\eList
\end{proposition}

\begin{proof} (i.) is obvious. If $\D_{11}(r)\not=0$ then, for
$|\O|>\big|\frac{\D_0(r,v_\theta)}
{v_\theta\D_{11}(r)}\big|$, $\Sign(S_3) =
\Sign(\O v_\theta)\Sign(\D_{11})$ and the other two statements follow from
Proposition \ref{p:RE-3}.
\end{proof}

Next, we investigate the leafwise-stability and instability of the
reduced equilibria of type RE3 with given $r$, as a function of $\O$
and $v_\theta$. Recall that for given $r$ and $\O$ there are two
branches of these equilibria in the plane $(v_\theta,\o_n)$, one with
$v_\theta>0$ and one with $v_\theta<0$, which are given by \eqref{omegatilden}
and are shown in Figure~2.

For any $\O\ge0$, the condition $S_3(r,v_\theta,\O)>0$ of
leafwise-stability (resp.  $S_3(r,v_\theta,\O)<0$ of
leafwise-instability) of the reduced equilibrium 
$\cP_3(r,v_\theta,\O)$ is:
\bList
\bull If $\D_{11}(r)=0$ 
\beq{stab-RE3-2}
  \D_{00}(r) + \D_{02}(r)v_\theta^2 + \D_{04}(r)v_\theta^4 >0
  \qquad\mathrm{\big(resp.}\ <0 \big) \,.
\eeq
\bull If $\D_{11}(r)\not=0$ 
\begin{eqnarray}
\label{stab-RE3-11}
  \tilde\O(r,v_\theta) \!\!\!\!&<&\!\!\!\! \O 
  \qquad\mathrm{\big(resp.}\ \tilde\O(r,v_\theta)> \O\big)
  \qquad \mathrm{if}\quad  v_\theta\D_{11}(r)>0
  \\
\label{stab-RE3-12}
  \tilde\O(r,v_\theta) \!\!\!\!&>&\!\!\!\! \O
  \qquad\mathrm{\big(resp.}\ \tilde\O(r,v_\theta) <\O\big)
  \qquad \mathrm{if}\quad  v_\theta\D_{11}(r)<0
\end{eqnarray}
with
\beq{stab-RE3-3}
  \tilde\O(r,v_\theta)
  \,:=\, 
  -\frac{\D_{00}(r)}{\D_{11}(r)}\frac1{v_\theta}
  -\frac{\D_{02}(r)}{\D_{11}(r)}{v_\theta}
  -\frac{\D_{04}(r)}{\D_{11}(r)}{v_\theta^3} \,.
\eeq
\eList
When $\D_{11}(r)\not=0$, for $\O=0$ conditions \eqref{stab-RE3-11}
and \eqref{stab-RE3-12} coincide with \eqref{stab-RE3-2}. Thus,
\eqref{stab-RE3-2} can be regarded as the condition for
leafwise-stability or instability when $\O=0$. For the case $\O=0$, a
condition equivalent to \eqref{stab-RE3-2} is given by Routh
\cite{routh}, 
who however does not
study or apply it, and appears also in  \cite{hermans} and
\cite{zenkov}.

(In some of the computations below we prefer using \eqref{stab-RE3-11} and
\eqref{stab-RE3-12} also for $\O=0$).

\begin{proposition}
\label{p:RE3-11}
For any $r>0$ and any $\O\ge0$:
\bList
\item[i.] In each of the two branches of reduced equilibria of type RE3
in the plane $(v_\theta,\o_n)$ there are at most two zeroes of the
function $S_3$. These zeroes divide the branch in up to three connected
components, in each of which all reduced equilibria are either
leafwise-stable or leafwise-unstable
(and if there are three, the properties of leafwise-stability
alternate among them).

\item[ii.] In each branch, the reduced equilibria in the closest component to
$v_\theta=0$ are leafwise-stable if $f'(r)f''(r)>0$ and
leafwise-unstable if $f'(r)f''(r)<0$. 

\item[iii.] In each branch, the reduced equilibria in the farthest
component from $v_\theta=0$ are leafwise-stable if
$rf''(r)>F(r)^2f'(r)$ and leafwise-unstable if $rf''(r)<F(r)^2f'(r)$.
\eList
\end{proposition}

\begin{proof}
(i.) Fix $\O$ and $r$. If $\D_{11}(r)=0$ then
$S_3(r,v_\theta,\O)=\D_{0}(r,v_\theta)$ is an even polynomial in
$v_\theta$ and we may study it only for $v_\theta>0$. Since it
has degree four, it has at most two positive roots. And if it has two
positive roots, none of them is an  extremal point. If
$\D_{11}(r)\not=0$, then the zeroes of $S_3(r,v_\theta,\O)$ are the
values of $v_\theta$ at which
$\tilde\O(r,v_\theta)=\O$. This is an odd function of $v_\theta$, and
again we may study it only for $v_\theta>0$. The positive zeroes of
$\tilde\O(r,v_\theta)$ are the positive roots of the even polynomial of
degree four $v_\theta\tilde\O(r,v_\theta)$.
Hence, they are at most two and $v_\theta\mapsto
\tilde\O(r,v_\theta)$ can have at most one extremal point on the
positive axis. 
It follows that, for $v_\theta>0$, its graph intersects
in at most two points any horizontal line. And if there are two
intersections, none of them is at an extrmal of $v_\theta\mapsto
\tilde\O(r,v_\theta,\O)$. (ii.) For small $|v_\theta|$, the sign of
$S_3(r,v_\theta,\O)$ is the same as that of $\D_{00}(r)$. (iii.) 
This follows from items ii. and iii. of Proposition \ref{p:RE-6}.
\end{proof}

We detail now a few situations, not with the purpose of being
exhaustive (which would require too many cases and
subcases, and can be done on a case by case basis) but with that of covering a
few typical situations and disclosing some general patterns. In
particular, we neglect almost all nongeneric cases. We define
$$
  \tilde\O_{m}(r)
  \;:=\;
  \inf_{v_\theta\not=0}|\tilde\O(r,v_\theta)|
$$

\vskip4mm\noindent
{\it Case 1: $f'(r)>0$, $f''(r)>0$.}
The three coefficients of the polynomial $\D_0(r,v_\theta)$ are all
positive (recall that $\mu<1$). Thus $\D_0(r,v_\theta)>0$ for all
$v_\theta\not=0$ and it follows that all RE3 reduced equilibria with this $r$
are leafwise-stable when $\O=0$. This has already been proved by
\cite{hermans}.

The situation for $\O>0$ depends on the sign of $\D_{11}(r)$. If
$\D_{11}(r)>0$, then the graph of $\tilde\O(r,v_\theta)$ is shown in
Figure 4.a, with $\tilde\O_{m}(r)$ finite and positive.
For $v_\theta>0$ the condition of leafwise-stability
is $\tilde \O(r,v_\theta)< \O$ and is satisfied by all reduced equilibria
because $\tilde \O(r,v_\theta)<0$ if $v_\theta>0$. For $v_\theta<0$,
$\cP_3(r,v_\theta,\O)$ is leafwise-stable if $\tilde \O(r,v_\theta)> \O$
and leafwise-unstable if $\tilde \O(r,v_\theta)<\O$. Thus, for
$\O<\tilde\O_{m}(r)$, all the reduced equilibria with $v_\theta<0$
are leafwise-stable. If instead $\O>\tilde\O_{m}(r)$ there are two values
$v_2<v_1<0$, which depend on $r$, such that $\cP_3(r,v_\theta,\O)$ is
leafwise-stable for $v_\theta<v_2$ and for $v_1<v_\theta<0$ and is
leafwise-unstable for $v_2<v_\theta<v_1$. See Figures 4.b and 4.c. 

As $\O\to\infty$, $v_2\to-\infty$ and $v_1\to0$ and the entire branch
of reduced equilibria with $v_\theta<0$ becomes leafwise-unstable, in
agreement with Proposition \ref{p:RE-6}. 

If $\D_{11}(r)<0$, then $\tilde\O$ has the opposite sign of that of the
case $\D_{11}(r)>0$; the resulting situation is depicted in Figures 4.b and 4.d.


\begin{figure}[h]
\begin{center}
{\small
{\scalebox{1.}{\includegraphics*{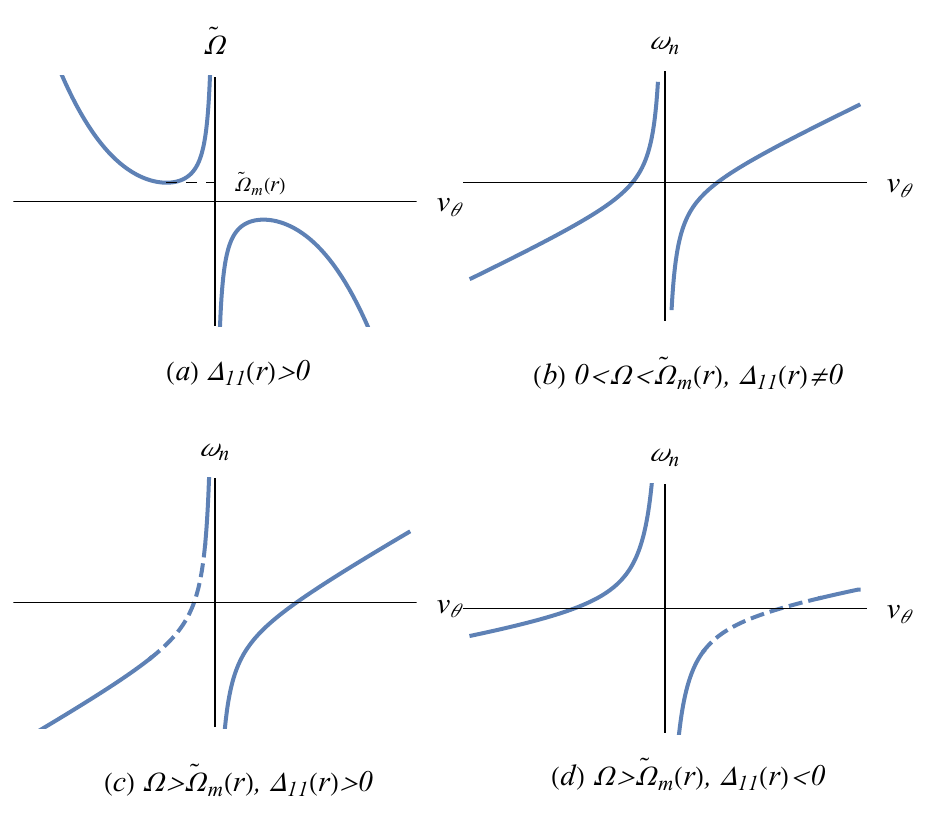}}}
}
\caption{\small Leafwise-stability of the two branches of reduced equilibria
of type RE3 in the plane $(v_\theta,\o_n)$ 
when $f'(r)>0$ and $f''(r)>0$ for various values of $\O$. 
(a) Graph of $\tilde\O(r,v_\theta)$ for
$\D_{11}(r)>0$. (b-d) Leafwise-stable (solid) and leafwise-unstable
(dotted) RE3 reduced equilibria for $\O>0$ and $\D_{11}(r)\not=0$.}
\end{center}
\end{figure}


We now study a few other cases. The analysis is similar to that of
Case 1, and we may limit ourselves to a few comments---mostly, to draw the
graph of the function $\tilde\O$. Instead of plotting
the bifurcation diagrams in the plane $(\o_n,v_\theta)$ we may
describe them by specifying the type and the order (left to right) of
the  components of leafwise-stability (``$S$'') and of
leafwise-instability (``$U$'') in each branch $v_\theta<0$ and
$v_\theta>0$. We write the resulting strings between brackets, with a
comma that separates the branch $v_\theta<0$ (first) from the branch
$v_\theta>0$. Thus, for instance, the bifurcation diagrams of
Figures 4.b-d are, respectively, of types $(S,S)$, $(SUS,S)$,
$(S,SUS)$. 

\vskip4mm\noindent
{\it Case 2: $f'(r)>0$, $f''(r)<0$.} In this case $\D_{00}(r)$ and
$\D_{11}(r)$ are negative,
$\D_{02}(r)$ is positive and $\D_{04}(r)$ may have any sign. 
If $\D_{04}(r)>0$ then the graph of $\tilde\O(r,v_\theta)$ is as in
Figure 5.a and the bifurcation diagram is of type $(SU,US)$. If 
$\D_{04}(r)<0$ there are two (generic) cases, depending on the sign of
the discriminant
$$
  D(r) := \D_{02}(r)^2 - 4|\D_{00}(r)\D_{04}(r)| \,.
$$
If $D(r)>0$ then the graph of $\tilde\O(r,v_\theta)$ is as in Figure
5.b, with $\tilde\O_m(r)$ finite and positive, and
the (generic) bifurcation diagrams are of type $(USU,USU)$ if
$0\le\O<\tilde\O_m(r)$ and of type $(USU,U)$ if $0>\tilde\O_m(r)$.
The graph of $\tilde\O(r,v_\theta)$ when $D(r)<0$ is as in Figure 5.c
and the (generic) bifurcation diagrams are of type $(U,U)$ if
$0\le\O<\tilde\O_m(r)$ and of type $(USU,U)$ if $\O>\tilde\O_m(r)$.


\begin{figure}[h]
\begin{center}
{\small
{\scalebox{1.}{\includegraphics*{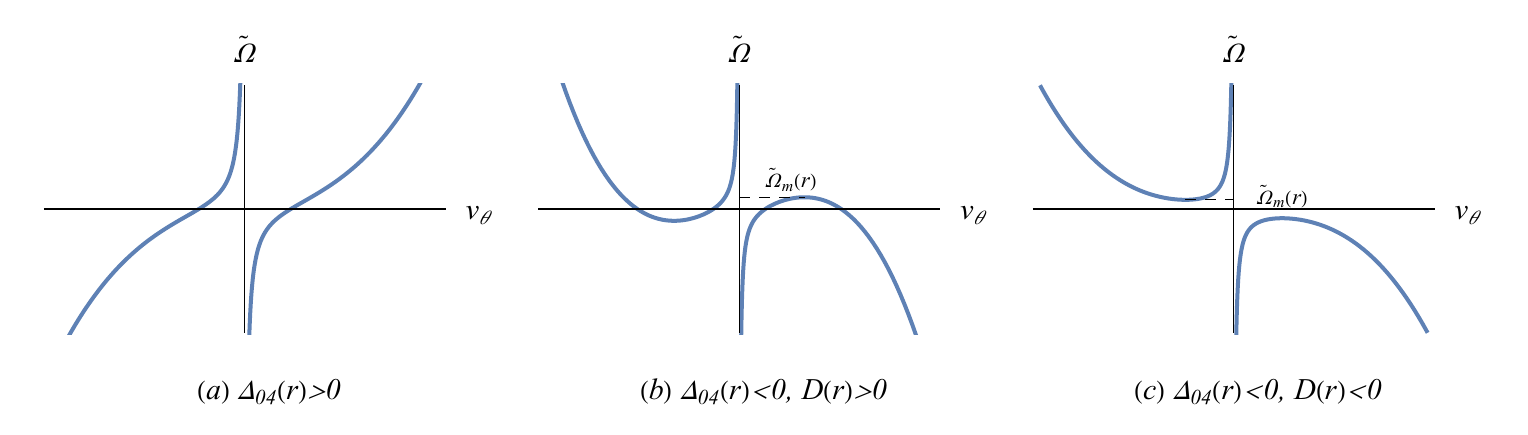}}}
}
\caption{\small Graphs of $\tilde\O(r,v_\theta)$ at fixed $r$ when $f'(r)>0$,
$f''(r)<0$: (a) $\D_{04}(r)>0$, (b) $\D_{04}(r)<0$ and $D(r)>0$, 
(c) $\D_{04}(r)<0$ and $D(r)<0$.}
\end{center}
\end{figure}


\vskip4mm
{\it Case 3: $f'(r)>0$, $f''(r)=0$.} This case is nongeneric, but it is
worth mentioning because it is the case of a cone, for which the
existence and stability of reduced equilibria has been investigated
in \cite{BIKM2019}.

In this case $\D_{00}(r)=0$, $\D_{02}(r)>0$, $\D_{04}(r)>0$ and
$\D_{11}(r)<0$. Thus $\D_0(r,v_\theta)>0$ for all $v_\theta\not=0$
and for $\O=0$ all RE3 reduced equilibria are leafwise-stable.
Moreover, $\tilde\O = \frac{\D_{02}}{|\D_{11}|}v_\theta
+\frac{\D_{04}}{|\D_{11}|}v_\theta^3$. It follows that, for $\O>0$,
all reduced equilibria with $v_\theta<0$ are  leafwise-stable while
those with $v_\theta>0$ are leafwise-stable for $v_\theta<v_1$ and
leafwise-unstable for $v_\theta>v_1$ with a certain $v_1>0$ that
depends on $r$ and $\O$ and goes to $0$ for $\O\to0$ and to $+\infty$
for $\O\to+\infty$. 

The bifurcation diagram is of type $(S,S)$ for $\O=0$ and of type
$(S,SU)$ for $\O>0$. 

\vskip4mm
{\it Case 4: $f'(r)<0$, $f''(r)<0$.} $\D_{00}(r)$ and $\D_{04}(r)$ are
positive, $\D_{02}(r)$ is negative and $\D_{11}(r)$ may have any sign. 
Let $D(r)$ be the discriminant defined in Case 2. 
\bList

\bull If $\D_{11}(r)>0$ and $D(r)>0$ then the graph of $\tilde\O$ is
as in Figure 5.b and the bifurcation diagram is of type $(SUS,SUS)$ if 
$0\le\O\le\tilde\O_{m}$ and of type $(SUS,S)$ if $\O>\tilde\O_{m}$.

\bull If $\D_{11}(r)>0$ and $D(r)<0$ then the graph of $\tilde\O$ is
as in Figure 5.c and the bifurcation diagram is of type $(S,S)$ if 
$0\le\O\le\tilde\O_{m}$ and of type $(SUS,S)$ if $\O>\tilde\O_{m}$.

\bull If $\D_{11}(r)<0$ and $D(r)>0$ then the graph of $\tilde\O$ is
the reflection about the $v_\theta$ axis of that shown 
in Figure 5.b. The bifurcation diagram is of type $(SUS,SUS)$ if 
$0\le\O\le\tilde\O_{m}$ and of type $(S,SUS)$ if $\O>\tilde\O_{m}$.

\bull If $\D_{11}(r)<0$ and $D(r)<0$ then the graph of $\tilde\O$ is
the reflection about the $v_\theta$ axis of that shown in Figure
5.c. The bifurcation diagram is of type $(S,S)$ if 
$0\le\O\le\tilde\O_{m}$ and of type $(S,SUS)$ if $\O>\tilde\O_{m}$.

\eList

\vskip4mm
Other cases can be studied similarly. 

\section{Example: the ball on an upward paraboloid}
\label{s:paraboloid}

\subsection{The parabolic surface. } We investigate now some aspects
of the dynamics for the parabolic profile
\beq{paraboloide}
   f = \frac12br^2
\eeq
with a constant $b>0$. This has two purposes. One is to prove
that, even if the profile is not superquadratic, all motions which do
not pass through the vertex are bounded, and hence generically
quasi-periodic, even for $\O\not=0$ (Proposition \ref{p:paraboloide};
this had been previously proven only for small values of $|\O|$, see
\cite{FS2015}). The other is to investigate, numerically, the
presence and number of (particulalry leafwise-unstable) reduced
equilibria on each level set of the map $J$. We give all expressions
in polar coordinates. Note that
$$
  f' = br \,,\qquad f''=b \,,\qquad F=\sqrt{1+b^2r^2} \,.
$$

\subsection{Reduced equilibria. } The system has only reduced equilibria of type
RE3, with
$$
  \tilde\o_n(r,v_\theta,\O) 
  =
  -\frac\g\mu \frac1{v_\theta} + \frac1{\mu b}v_\theta 
  -\O\Big(\frac1b+\frac1F\Big) \,,\qquad v_\theta\not=0 \,.
$$
The two branches they form are independent of $r$ if $\O=0$, but for
$\O>0$ they are shifted below by an amount which decreases with $r$
and varies between $\O(1+\frac1b)$ and $\frac\O b$.

All these reduced equilibria pertain to case 1 of Section
\ref{ss:stab-RE3}, with  $\D_{11}=-\g \mu b^2 r^3<0$. For $\O=0$ they
are all leafwise-stable and for $\O>0$ all those with negative
$v_\theta$ are leafwise-stable. We thus focus on the
reduced equilibria with $\O>0$ and $v_\theta>0$. 

For $v_\theta>0$, the possible situations are those of Figures 4.b
and 4.d. The function $\tilde\O$ is given by
$$
  \tilde\O(r,v_\theta) = 
  \frac\g{b\mu r^2 v_\theta} 
  + \frac2{\mu}\Big(\frac1{b^2r^2}+1\Big) v_\theta
  + \frac1{b\g\mu}\Big(\frac1{b^2r^2}+2+\mu b^2r^2\Big)v_\theta^3 
$$
and some of its level curves (with values increasing from top to
bottom) are shown in Figure 7 for three different values of the
parameter $b$.
For each $\bar r>0$, the level curve  $\tilde\O(\bar
r,v_\theta)=\tilde\O_{m}(\bar r)$ is the one tangent to the
(horizontal) line $r=\bar r$. Therefore, the function
$r\mapsto\tilde\O_{m}(r)$ is a strictly decreasing function which
tends to $+\infty$ for $r\to0$ and to $0$ for $r\to+\infty$ and its
graph resembles that of a branch of a hyperbola. Its inverse
$\O\mapsto\tilde r_{m}(\O)$, which gives the $r$-coordinate of the
minimum of the level curves of $\tilde\O$, has these same
properties. We stress that, for each $\O>0$, $\tilde r_m(\O)>0$.

\begin{figure}[h]
\begin{center}
{\small
{\scalebox{1.}{\includegraphics*{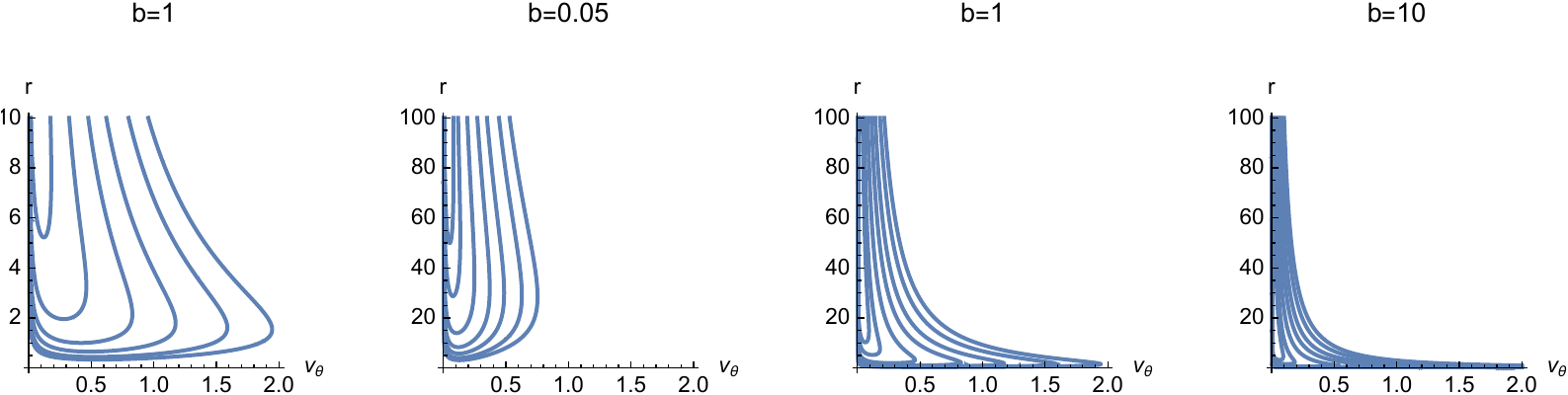}}}
}
\caption{\small Level curves of $\tilde\O$  for the
parabolic profile \eqref{paraboloide}.  Each panel shows (from bottom
to top) the level curves $\tilde\O=100,60,30,15,6,2,1$ for the
indicated values of $b$. Figure 7.a is an enlargement of Figure 7.c.}
\end{center}
\end{figure}


This provides the following picture for the stability of the reduced
equilibria $\cP_3(r,v_\theta,\O)$ with $v_\theta>0$. For each $\O>0$,
they are all leafwise-stable if $r<\tilde r_m(\O)$.
For $r>\tilde r_m(\O)$ there are the three intervals S-U-S of values of
$v_\theta$ as in Figure 4.d. As $r$ increases, the first ``S''
interval, the one closest to $v_\theta=0$, 
becomes extremely narrow while the amplitude of the middle ``U''
interval reaches a maximum and then goes (slowly) to zero as
$r\to+\infty$. We stress that all reduced equilibria become stable
for $v_\theta$ large enough.

Concerning the dependence on the parameters, Figure 6 indicates that
as $b$ increases, namely, as the paraboloid becomes steeper, the
amplitude of the U-interval decreases at small $r$ but increases at
large $r$. We mention that increasing $\g$ expands the instability
region at all $r$, while increasing $\mu$ expands it at small $r$ and
seems to have little effect at large $r$. 

\subsection{The $J$-restricted reduced systems. } In order to understand the
dynamics we investigate now the $J^\circ$-restricted systems. 
For the parabolic profile the integration of the differential
equations that give the two first integrals $J_1$ and $J_2$ can be
done explicitly. Expressed as functions of $r$ instead of $p_1$, the
solutions of equations \eqref{naode1} and \eqref{naode2} are
$$
  \overline U = 
  \left( \begin{matrix}
    c & \frac{\sqrt\mu}b s \cr
   \frac b{\sqrt\mu}s & c
  \end{matrix}  \right)
  \,,\qquad 
  \overline u = 
  \left(\begin{matrix}
    \frac1{(4-\mu)b^2} 
    \big( (4-3\mu)(c-1)+(4b-(b+1)\mu)\sqrt\mu\,s + 2 \mu b^2r^2 \big)
  \cr
    \frac1{(4-\mu)b} 
    \big(
    (4b-(b+1)\mu)c + \frac{4-3\mu}{\sqrt\mu}s
     - b \frac{4-\mu}{F}+\mu F^2
    \big)
  \end{matrix}
  \right)
$$
where
$$
  c(r) := \cosh\Big(\frac{\sqrt\mu}2\log F(r) \Big) 
    = \frac{F(r)^{\sqrt\mu}+1}{2F(r)^{\sqrt\mu/2}}
  \,,\qquad
  s(r) := \sinh\Big(\frac{\sqrt\mu}2\log F(r) \Big) 
    = \frac{F(r)^{\sqrt\mu}-1}{2F(r)^{\sqrt\mu/2}} \,.
$$
The effective potential $V_j$ is given by \eqref{Vj} with
$$
  \left(\begin{matrix}
  \overline p_{3,j}(r) \cr \overline p_{4,j}(r) 
  \end{matrix} \right)
  =
  \bar U(r)j+\O\bar u(r) 
$$
see \eqref{tildep3p4}. 

\begin{proposition} 
\label{p:paraboloide}
If $f(r)=\frac12br^2$, $b>0$, then for any
$j_1\not=0$ and $\O\ge0$, $V_\Oj(r)$ goes to $+\infty$ for $r\to0^+$ and
for $r\to+\infty$.  
\end{proposition}

\begin{proof}
Since $\overline p_3(0,j)=j_1$ and $\overline p_4(0,j)=j_2$, for
$r\to0^+$ the function $V_\Oj$ is asymptotic to
$\frac{j_1^2}{2r^2}+\frac12\mu j_2^2+\O(\mu j_2-j_1)-\frac12\mu\O^2$.
For $r\to+\infty$, $c(r)$ and $s(r)$ are both asymptotic to
$r^{\sqrt\mu/2}$ and the same is true for the matrix $\overline U(r)$.
Instead, $\overline u(r)$ is asymptotic to $r^2$. Thus, both $\overline
p_{3j}$ and $\overline p_{4j}$ are asymptotic to~$r^2$. This implies
that, if $\O>0$, then $V_\Oj$ is asymptotic for $r\to+\infty$ to
$\frac12\mu\overline p_{4j}(r)^2$ and hence to $r^4$.
\end{proof}

This implies that, for all $j_1\not=0$, the dynamics of the reduced
system is periodic (and hence that of the unreduced one is
quasi-periodic) except for the equilibria and the motions asymptotic
to and from the unstable ones. We do not investigate here motions in
the level set $j_1=0$ because it contains the vertex.

\subsection{The equilibria of the $J$-restricted reduced systems. } 
Proposition \ref{p:paraboloide} implies that, for any $j_1\not=0$
and $\O\ge0$, the effective potential has at least one minimum, and
hence the restriction of the reduced system to $M_2^j$ has at least
one stable equilibrium. In fact, since $V_\Oj$ is a real analytic nonconstant
function, its minima are all isolated and, since the system
has one degree of freedom, they are the only stable equilibrium
configurations. Generically, there is obviously an odd number of
equilibria on each $M^2_j$, but their exact number---and the numbers
of the stable and unstable ones---is of special interest because gives
global information on the dynamics in $M_2^j$.

We already know that, when $\O=0$, all reduced equilibria are
leafwise-stable. This implies that, for each $j\not=0$, $V_j$ has a
single critical point, which is a minimum, and the reduced system in
$M_2^j$ has only one equilibrium. Figure 8.a shows, for
a typical choice of the values of the parameters $\mu$, $b$, $j$,
the value of the $r$-coordinate of the reduced equilibrium on
$M^2_j$ as a function of $j=(j_1,j_2)$. This is a single valued
surface. At fixed $j_2$, the $r$-coordinate of the reduced
equilibrium tends to a constant value when $|j_1|\to\infty$ and there
is a single maximum of $r$, which goes to $+\infty$ when $|j_2|\to
\infty$. Not surprisingly, when $j_1\to0$ the coordinate the reduced
equilibrium tends to the vertex ($r\to0$). Note the symmetry of the
surface $S$ under reflections of $(j_1,j_2)$.

In order to determine the number of unstable equilibria for $\O\not=0$ we
resorted to a numerical analysis, whose results are illustrated by
Figures 7.b-d. As soon as $\O\not=0$, two (or exceptionally, at the
bifurcations, one) other reduced equilibria are created for $j=(j_1,j_2)$
in about half of the $(j_1,j_2)$-plane, one of which is leafwise-unstable and
the other (if present) is leafwise-stable. The figures show the
equilibria surface for different values of $\O$ and in different
ranges of $j_1,j_2$. Even though the figures cannot show it clearly,
the shape of the surface is similar for all values of $\O$ but, as
one sees observing that the figures have different scales, 
as $\O\to 0$ the two
additional equilibria go to infinity in $r$ and/or $v_\theta$. 



\begin{figure}[h]
\begin{center}
{\small
{\scalebox{.7}{\includegraphics*{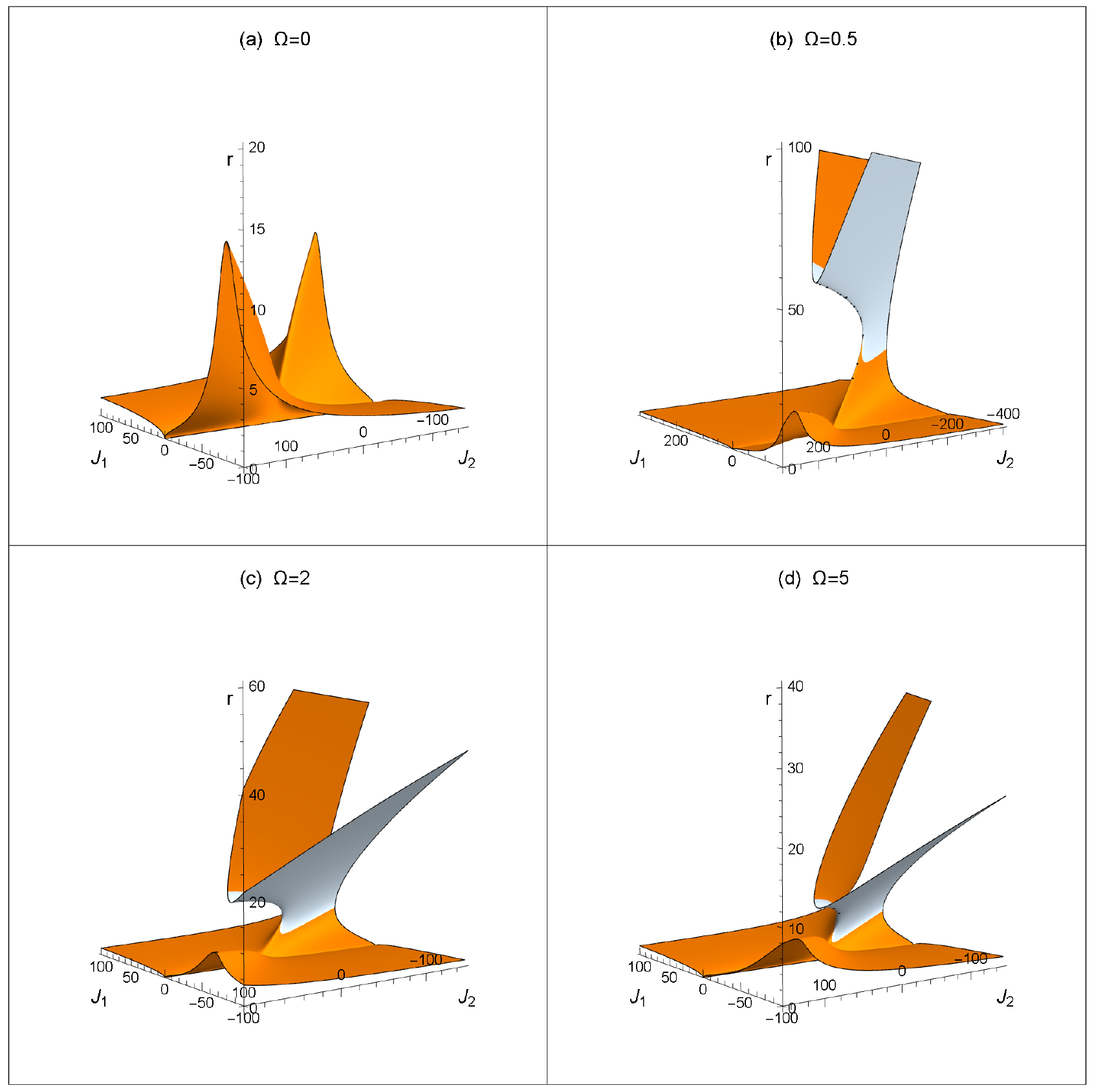}}}
}
\caption{\small The value of the $r$ coordinates at the
reduced equilibria for the parabolic profile \eqref{paraboloide}
($b=\gamma=1$, $\mu=2/7$)}
\end{center}
\end{figure}


\section{Conclusions. } We have provided a general analysis of the
dynamics of a heavy dynamically symmetric ball that rolls without
sliding on a uniformly rotating surface of revolution. Even though
this study has clarified a number of aspects of this class of systems,
some questions remain open.

1. The possibility and the properties of motions through---or
asymptotic to---the vertex have not been studied. The possibility of
motions in which the point of contact tends to the vertex and (some
component of) the angular velocity grows unbounded is not ruled out by
our analysis and should be invetigated. One natural possibility is to
analyse these motions in the five-dimensional $\SO(3)$-reduced
system.  

2. When $\O\not=0$, we have only proven the boundedness of motions
under the hypothesis that the profile of the surface goes superquadratically
to $+\infty$ at infinity. We have proven this fact using the
compactness of the level sets of the moving energy. However, as
pointed out in section \ref{ss:superquadratic}, what is necessary is
the compactness of all the level sets of the map $(E,J)$ which, as the
example of the (upward) paraboloid of section \ref{s:paraboloid}
shows, might be satisfied under the assumption alone of coercitivity of
the profile. A general study of this question might require a careful
analysis of the asymptotic properties of the functions $J_1$ and
$J_2$ defined by the differential equations \eqref{p3p4}.

3. When $\O=0$, if the profile goes asymptotically to $-\infty$, or
to a constant, then there are certainly unbounded motions, in which
the ball goes to infinity. Even though some particular statements are
made by Routh in \cite{routh}, a characterization of the initial
conditions which lead to bounded or unbounded motions is essentially
missing.

4. In connection with point 3., we remark that the example of the
ball that rolls on a horizontal plane suggests that the rotation of
the surface may have a `stabilizing' effect on the dynamics. In fact,
in all motions but the equilibria the ball runs away to infinity if
$\O=0$, but as soon as $\O\not=0$ the ball moves on circles! 
Preliminary investigations show that such a stabilizing effect of the
rotation is present in other profiles, e.g. in the downward paraboloid
and cone, and we conjecture that, as soon as $\O\not=0$, all motions
in {\it any} profile are bounded (with the possible exception of
those asymptotic to the vertex).

5. Also the (local and global) structure of the foliation by the
invariant tori (in integrable cases) is still not studied. This study
would require some comprehension of the frequencies of motions. Some
results on this, for the case of a corcive profile and $\O=0$
are given by \cite{hermans}.

\section{Appendix: The equations of motion}

\subsection{The nonholonomic equations of motion in quasi-velocities with the reaction
forces. }
The equations of motion of mechanical systems subject to nonholonomic
constraints can be written in several ways. Particularly when the
configuration space involves a Lie group it is customary to employ
a technique originally developed by Poincar\'e for holonomic systems
\cite{poincare}, which is based on the use of coordinates and
quasi-velocities---namely linear combinations of the velocities. 
For instance, for rigid bodies this
allows to use the components of the angular velocity (with respect to
a fixed or moving frame) instead of the velocities of the Euler angles
or other local coordinates on $SO(3)$. The nonholonomic case was first
considered by Hamel \cite{hamel}.

However, in Hamel's approach the quasi-velocities are chosen so that the
nonholonomic constraint is given as zero of some of them. This leads
to a set of equations on the constraint manifold---Hamel
equations---in which the reaction forces are
not explicitly identified. In our opinion, instead, the explicit
consideration of the reaction forces is under several respects
important, e.g. in determining the conservation of momenta and energy
\cite{FS2015,FS2016,FNS2018}. 

We thus derive here a form of the equations of motion of
nonholonomic systems that employs quasi-velocities and contains, in an
explicit way, the reaction forces. Specifically, we write these
equations as the restriction to the nonholonomic constraint manifold
$M\subset TQ$ of a set of equations in the tangent bundle of the
configuration manifold $Q$ that leave $M$ invariant (namely, as a
vector field which is tangent to $M$). This is a generalization of an
analogous form of the equations that uses Lagrangian coordinates and
velocities, which is our starting point and for which we refer to
\cite{FS2015}.

We consider a nonholonomic system $(Q,\cL,\cM)$ with an
$n$-dimensional configuration manifold $Q$, a mechanical Lagrangian
$\cL:TQ\to\bR{}$, and an affine distribution $\cM$ on $Q$ with
constant rank that describes the nonholonomic constraint. More
specifically:
\bList
\item[i.] By a mechanical Lagrangian we mean a function of the form
$\cL=\cL_2+\cL_1+\cL_0$, where $\cL_2$ is a Riemannian metric on $Q$,
$\cL_1$ is a function whose restriction to each fiber of $TQ$ is
linear, and $\cL_0$ is a basic function, hence constant on the fibers
of $TQ$. 

\item[ii.] We write the affine distribution as $\cM=\xi+\cD$, with
$\cD$ a non-integrable distribution on $Q$ of constant rank $r$,
$1<r<n$, and $\xi$ a vector field on $Q$.  Clearly, the vector field
$\xi$ is defined up to a section of~$\cD$. We denote by $M$ the
$(n+r)$-dimensional subbundle of $TQ$ formed by the fibers of $\cM$.

\item[iii.] Lastly, we assume that the nonholonomic constraint is ideal,
namely, that the reaction forces it exerts satisfy d'Alembert
principle, see \cite{agostinelli,FS2015} for details. 
\eList
It is well known that, under these hypotheses, there is a unique function
$R_{\cL,M}:M\to\cD^\circ$ with the property that the restriction to $M$
of Lagrange equations with the reaction forces,
\begin{equation}
\label{EqLagrWithRF}
   [\cL]\big|_{M}  = R_{\cL,M} \,,
\end{equation}
defines a vector field on $M$, and hence a dynamical system on $M$
\cite{agostinelli,FS2015}.
Here, $[\cL]$ is the usual Euler-Lagrange operator. The expression of
these equations using lifted coordinates $(q,\dot q)$ in $TQ$ is
given in \cite{FS2015}, and can be recovered as a particular case of
the present treatment.

Consider now a set of local coordinates $q:Q_U\to U$ defined in an open set
$Q_U \subseteq Q$ and taking values in an open set $U$ of $\bR n$. We
call `lifted coordinates' the coordinates $(q,\dot q)$ in $TQ_U$.
Consider a smooth function $B:U\to GL(n)$. Then, the change of
coordinates
$$
   (q,\dot q) \mapsto \big(q,B(q)\dot q\big) =:(q,v)
$$
defines a new set of bundle-like coordinates in $TQ_U$. The
expression of the Euler-Lagrange operator $[\cL]$ in these
coordinates is well known (Lagrange-Poincar\'e equations
\cite{poincare}), and we need only compute $R_{\cL,M}$. 

The local representative $L$ of the Lagrangian $\cL$ has the form
$L=L_2+L_1+L_0$ with $L_0$ independent of the $v$'s, $L_1$ linear in the
$v$'s, and $L_2(q,v)=\frac 12 v\cdot A(q)v$ with 
$$
   A= \dder Lvv
$$
a positive definite matrix that depends only on $q$. The fibers of the
distribution $\cD$ based in $Q_U$ can be represented as the kernel of
a $q$-dependent $(n-r)\times n$ matrix $S(q)$ of rank $n-r$: the
fiber of $\cM$ based at the point of $Q_U$ of coordinates $q$ is
given by the equation
$$
   S(q)v+s(q)=0 \,,
$$
where $q\mapsto s(q)\in\bR n$ is a a smooth map that depends on the
vector field $\xi$ (specifically, $s(q)=-S(q)\xi^\mathrm{loc}(q)$ if
$\xi|_{Q_U}=\sum_i\xi^\mathrm{loc}_i\partial_{q_i}$). The image of $M\cap
TQ_U$ under the coordinate map $(q,v)$ is the $(n+r)$-dimensional
submanifold
$$
  M_U :=  \big\{ (q,v)\in U\times\bR n\,:\; S(q)v+s(q)=0\big\} 
$$
of $U\times \bR n$. 

Define now maps $\ell:U\times\bR n\to\bR n$,
$\sigma:U\times\bR n\to\bR{n-r}$ and $R:U\times\bR n\to\bR n$ as
follows: $\ell(q,v)$ has components\footnote{We understand
summation over the repeated indexes $i,j,h,k,l=1,\ldots,n$.}
\begin{equation}
\label{eq:ell}
  \ell_i = \dder L {v_i}{q_j} \,B^{-1}_{jh}\,v_h
           + \g_{ijh} \,\der L{v_j} \,v_h
           - B^{-T}_{ij}\der L{q_j} 
  \,,\qquad i=1,\ldots,n\,,
\end{equation}
where 
$\g_{ijh} = B^{-T}_{ik}\big(\der {B^T_{kj}}{q_l} -
\der{B^T_{lj}}{q_k}\big) B^{-1}_{lh}$ are the so-called ``transpositional
symbols'', $\sigma(q,v)\in\bR k$ has components
\begin{equation}
\label{eq:sigmatilde}
  \sigma_a = \Big( \der{S_{ai}}{q_j}v_i 
                    +
                   \der{s_a}{q_j}\Big) B^{-1}_{jh}v_h 
  \,,\qquad a=1,\ldots,n-r \,,
\end{equation}
and 
\begin{equation}
\label{eq:Rtilde}
    R = S^T (SA^{-1}S^T)^{-1}    (SA^{-1}\ell-\sigma) \,.
\end{equation} 

\begin{proposition}
\label{p:PoiHam}
The representative of equation \eqref{EqLagrWithRF} in the coordinates
$(q,v)$ is the restriction to $M_U$ of the equation
\begin{equation}
\label{eq:ph_eqs}
   \dot q = B(q)^{-1} v \,,\qquad A(q)\,\dot v+ \ell(q,v) = R(q,v) 
\end{equation}
in $U\times\bR n$.
\end{proposition}

\begin{proof}
The representative $\tilde L$ of $\cL$
in the coordinates $(q,\dot q)$ is $\tilde L(q,\dot q) = L(q,B(q)\dot
q)$ and so $\tilde A:=\dder{\tilde L}{\dot q}{\dot q}= B^TAB$.
The constraint manifold in lifted coordinates is given by $\tilde S(q)\dot
q+s(q)=0$ with $\tilde S = SB$. The equations of motion in lifted
coordinates are known \cite{FS2015} to be the
restriction to $\tilde M_U = \{(q,\dot q)\in U\times \bR n \,:\;
\tilde S(q)\dot q + s(q)=0\big\}$ of the equation $\tilde A\ddot q +
\tilde\ell=\tilde R$, where $\tilde\ell$, $\tilde\sigma$ and $\tilde R$ are
defined by formulas \eqref{eq:ell}--\eqref{eq:Rtilde} with $L$, $A$ and
$S$ replaced respectively by $\tilde L$, $\tilde A$ and $\tilde S$,
$v$ replaced by $\dot q$, and $B$ replaced by the unit matrix (hence all $\g_{ijh}$
vanish). A computation gives
\begin{eqnarray*}
  \tilde A(q)\ddot q
  \!\!\!&=&\!\!\! 
  B(q)^T\big[ A(q)\dot v - A(q)\dot B(q,\dot q)\dot q \big]
  \\
  \tilde \ell(q,\dot q)
  \!\!\!&=&\!\!\! 
  B(q)^T\big[\ell(q,v) + A(q)\dot B(q,\dot q)\dot q\big]
  \\
  \tilde \sigma(q,\dot q)
  \!\!\!&=&\!\!\! 
  \sigma(q,v) + S(q)\dot B(q,\dot q)\dot q
\end{eqnarray*}
where $v$ stands for $B(q)\dot q$ and $\dot B(q,\dot q)$ is the matrix
with entries $\dot B_{ij}=\der{B_{ih}}{\dot q_j}\dot q_h$.
Thus $\tilde A \ddot q+\tilde \ell-\tilde R =
B^T\big(A\dot v+\ell-R)$. 
\end{proof} 

\subsection{The (reduced) equations of motion of our system. }

We write now the equations of motion of the ball on the rotating
surface of revolution considered in this paper and of its
$SO(3)\times\SO(2)$-reduction. 

In order to facilitate the reduction under the $SO(2)$-action we
restrict at first to the subset $M_8^\circ$ of $M_8=\bR2\times
SO(3)\times\bR2\times \bR{} \ni (x,\cR,\dot x, \o_\z)$ where $x\not=0$
and use polar coordinates $(r,\theta)\in\bR{}_+\times S^1$ 
in its factor $\bR2\setminus\{0\}\ni x$, with $x_1= r\cos\theta$,
$x_2=r\sin\theta$. Correspondingly, we restrict the holonomic system
to the submanifold $M_{10}^\circ$ of $M_{10}=\bR2\times SO(3)\times\bR2\times
\bR3 \ni (x,\cR,\dot x, \o)$ where $x\not=0$ and here too we use polar
coordinates in the factor $\bR2\times\{0\}$, thus working in
$$
   \widehat{M}_8^\circ =
   \reali_+\times S^1 \times \SO(3)\times\bR{}\times \bR{}\times \bR{}
    \ni (r,\theta,\cR,v_r,v_\theta , \o_\z) \,.
$$
Furthermore, in
$M_{10}^\circ$ we use the quasi-velocities
$$
  v = (v_r,v_\theta,\o_\x,\o_\y,\o_\z) \,,
$$
thus identifying $M_{10}^\circ$ with  $\bR{}_+\times S^1\times
SO(3)\times\bR{}\times \bR{}\times \bR{3}$.
The representative of the Lagrangian \eqref{eq:Lagrangian}
of the system is
$$
  L(r,\theta,\cR,v_r,v_\theta,\o_\z) 
  =
  \frac12F(r)^2v_r^2 + \frac12r^2v_\theta^2 
  + \frac12 k \|\o\|^2 - \hat g f(r)
$$
and the nonholonomic constraint \eqref{vinc2} becomes
$\o_\x=\tilde\o_\x(r,\theta,v_r,v_\theta,\o_\z)$,
$\o_\y=\tilde\o_\y(r,\theta,v_r,v_\theta,\o_\z)$ 
with
\begin{equation}
\label{ParVinc}
\begin{aligned}
  \tilde \o_\x &=
  (\Omega-\o_\z)f' \cos\theta + 
     (\Omega r\cos\theta -v_r\sin\theta-r v_\theta\cos\theta)F
  \\
  \tilde \o_\y &=
  (\Omega-\o_\z)f'\sin\theta + 
      (\Omega r\sin\theta +v_r\cos\theta-r v_\theta\sin\theta)
      F
   \,.
\end{aligned}
\end{equation}
Accordingly, we identify $M_8^\circ$ with
$$
  \widehat{M}_8^\circ := 
  \bR{}_+\times S^1\times SO(3)\times\bR{}\times \bR{}\times \bR{}
  \ni (r,\theta,\cR,v_r, v_\theta, \o_\z) \,.
$$
In this identification, $SO(3)\times SO(2)$ acts on the factor
$\bR{}_+\times S^1$ by translations
in $S^1$ and $M_8^\circ/SO(3)\times SO(2)$ can be identified with 
$$
  \widehat{M}_4^\circ := 
  \bR{}_+\times \bR{}\times \bR{}\times \bR{}
  \ni (r,v_r, v_\theta, \o_\z) \,.
$$

\begin{proposition}
\label{p:EqMoto}
(i) The equations of motion in $\widehat{M}_8^\circ$ are
\begin{eqnarray*}
  \dot r \ugarr v_r 
  \\
  \dot \theta \ugarr v_\theta
  \\
  \dot \cR \!\!\!&=&\!\!\!
     \cR^T \big(\tilde\o_\x,\tilde\o_\y,\o_\z
         \big)^T 
  \\
  \dot v_r
  \ugarr
  -\g f'F^{-2} - f'f''F^{-2}v_r^2 + r(1+\mu f'^2)F^{-2}v_\theta^2  
  +\mu f'F {-v}_\theta \o_\z
  - \Omega \mu(r+f'F^{-1})v_\theta
  \\
  \dot v_\theta
  \ugarr 
  -\frac{v_r}r\Big[
     \big(2+\mu r f'f''F^{-2}\big)v_\theta
     + \mu f''F^{-1} \o_\z 
     - \Omega\mu(1 +f''F^{-1} + r f'f''F^{-2})
   \Big]
  \\
  \dot\o_\z
  \ugarr
  -v_r\frac1{1+k} f'F^{-1}\Big[
     rf'f''F^{-2}v_\theta 
     + f'' F^{-1} \o_\z
     - \Omega\big(1+ f''F^{-1} + rf'f''F^{-2}\big)
    \Big]
\end{eqnarray*}
with  $\mu=\frac k{k+1}$, $\g=\frac{\hat g}{k+1}=\frac g{(k+1)a}$
and $F=(1+f'^2)^{1/2}$ (see \eqref{simboli}).

(ii) The equations of motion of the reduced system in $\widehat{M}_4^\circ$ are 
given by the first and the last three equations in (i).
\end{proposition}

\begin{proof} (i) In order to invoke Proposition \ref{p:EqMoto} we need to
introduce local coordinates $\a=(\a_1,\a_2,\a_3)$ in $SO(3)$. Due
to the $SO(3)$-symmetry, the choice of these coordinates is
irrelevant, but in order to be able to consider a single chart it is
convenient to choose them so that their domain is open and dense in
$SO(3)$. For instance, we could use three Euler angles.

The function $\ell$ can be computed without using its expression
\eqref{eq:ell} because Lagrange equations for the
holonomic system of Lagrangian $\cL$ are $A\ddot q +\ell=0$, with
$A=\mathrm{diag}(F^2,r^2,k,k,k)$.
Since $v_r$ and $v_\theta$ are velocities, detailing the
corresponding Lagrange equations gives the first two components of
$\ell$. Since $\o$ is a first integral of the holonomic system, 
the last three components of $\ell$ are all zero. Explicitly,
and using $FF'=f'f''$,
$$
  \ell = 
  \big( f'f'' v_r^2 - r v_\theta^2 + \hat g f', 
        2r v_r v_\theta ,  0,  0,  0  \big) \,.
$$
Next, the matrix that gives the quasi-velocities
$v=(v_r,v_\theta,\o_\x,\o_\y,\o_\z)$ is
$$
  B(r,\theta,\a) = \mathrm{diag}\big(1,1,b(\a)\big)
$$
with a certain $3\times3$ invertible matrix $b(\a)$ whose
expression is irrelevant.  The nonholonomic constraint \eqref{vinc2}
can be written as $S(r,\theta)v+s(r,\theta)=0$ with
$$
  S(r,\theta) = 
  \left(\begin{matrix}
     F \cos\theta & -r F\sin\theta & 0 & -1 &-f'\sin\theta 
     \cr
     F \sin\theta & r F\cos\theta & 1 &0  &f'\cos\theta 
     \end{matrix} \right) 
  \,,\qquad
  s(r,\theta) = \Omega
  \left(\begin{matrix}
     (r F+f')\sin\theta 
     \cr
     -(r F+f')\cos\theta
     \end{matrix} \right) \,.
$$
A direct computation gives
$$
  (SAS^T)^{-1} = \mu \cF^2
  \left( \begin{matrix}
       1+f'^2(\cos\theta)^2  &  f'^2 \cos\theta\,\sin\theta \cr
       f'^2 \cos\theta\,\sin\theta & 1+f'^2(\sin\theta)^2  
  \end{matrix} \right) \,.
$$
Since $S$ and $s$ are independent of the $\a$'s, the sum over the
index $j$ in the expression \eqref{eq:sigmatilde} of $\s$ reduces to
$j=1,2$. Since $B^{-1}_{jh}=\delta_{jh}$ for $j=1,2$ and
$h=1,\ldots,5$, we have
$$
  \s_a =
  \der{S_{ah}}r v_r v_h + \der{s_a}r v_r
  +
  \der{S_{ah}}\theta v_\theta v_h + \der{s_a}\theta v_\theta 
  \,,\qquad a = 1,2 \,,
$$
and in fact, since the third and fourth component of $S$ are constant,
all sums over the index $h$ restrict to $h=1,2,5$. (This implies that
$\s$, as all other terms, is independent of $\o_\x$ and $\o_\y$; this
will make the restriction to the constraint manifold trivial). Putting the
various terms together, and using again the identity $F'=f'f''/F$, we
eventually find from \eqref{eq:Rtilde}
\begin{equation}
\label{R}
  R =  \mu
  \left(\begin{matrix}
     \hat g f' + (r f' v_\theta^2 + F\o_\z v_\theta) f'
     \\
     -(r f'v_r v_\theta + F\o_\z v_\theta)r f''F^{-2} 
     \\
     *
     \\
     *
     \\ 
     -(r f'F^{-1}  v_rv_\theta + v_r\o_\z )f' f'' F^{-2}
  \end{matrix}\right)
  +
  \Omega \mu 
  \left(\begin{matrix}
     -(r F + f')Fv_\theta
     \\
     \big(1+f''F^{-1}+r f'f''F^{-2} \big)r v_r
     \\
     *
     \\
     *
     \\ 
     \big(1+f''F^{-1} +r f' f'' F^{-2} ) f' F^{-1} v_r
\end{matrix}\right) 
\end{equation}
where the third and fourth components are not detailed because they
will be eliminated by the restriction to the constraint manifold.

In conclusion, the equations of motion \eqref{eq:ph_eqs} in the
10-dimensional manifold $\widehat{M}_{10}^\circ$ are given by the equations of
Proposition \ref{p:EqMoto} with the third one replaced by
$$
  \dot\a = b(\a)^{-1}\o 
$$ 
and with the two equations for $\dot\o_\x$ and $\dot \o_\y$ added.
Obviously, the equation for $\dot\a$ is the representative in the
chosen coordinates of the equation $\dot \cR=\cR^T\o$. Since we have
assumed that the domain of the coordinates $\a$ is dense in $SO(3)$,
by continuity we may conclude that the equations of motion
\eqref{eq:ph_eqs} are given by this set of equations with that for
$\dot\a$ replaced by $ \dot \cR = \cR^{-1}\o$. The restriction to
$\widehat{M}_8^\circ$ is performed by ignoring the equations for $\dot\o_\x$ and
$\dot\o_\y$ and replacing $\o_\x$ and $\o_\y$ with $\tilde\o_\x$ and
$\tilde\o_\y$ wherever they appear in the others (namely, in the
equation for $\dot\cR$).

(ii) This is obvious.
\end{proof}

We can now deduce the reduced equations \eqref{X}. In the subset $M_4^\circ$ of
the phase space we may use as (global) coordinates the four functions
$p_1,p_2,p_3,p_4$ as in \eqref{PolInv}, whose expression in polar
coordinates is
$$
  p_1 = \frac{r^2}2 \,,\qquad 
  p_2 = r v_r \,,\qquad
  p_3 = r^2 v_\theta \,,\qquad
  p_4 = -\big(F \o_\z+rf'v_\theta\big) 
        + \Omega \big( r +f'F^{-1} \big)f' \,,
$$
where the latter is obtained by observing that, in $\widehat{M}_8^\circ$, 
$\o\cdot n=(f'\tilde \o_\x\cos\theta+f'\tilde\o_\y \sin\theta -
\o_\z)F^{-1}$ with $\tilde\o_\x$ and $\tilde\o_\y$ as in \eqref{ParVinc}.
The inverse change of coordinates, which uses $\psi$ and $\cF$ instead
of $f$ and $F$, is given by
$$
  r = \sqrt{2p_1} \,,\qquad 
  v_r = \frac{p_2}{\sqrt{2p_1}}  \,,\qquad
  v_\theta = \frac{p_3}{2p_1} \,,\qquad
  \o_\z = -\big(p_4+\psi'p_3) \cF
         + \Omega \big(1+\psi'\cF\big)2p_1\psi'\cF\,.
$$
From here, a computation shows that, in $M_4^\circ$, the first and
the last three of the equations of Proposition \ref{p:EqMoto} become the
four equations $\dot p_i=X_i|_{M_4^\circ}$, $i=1,\ldots,4$, with
$X_1,X_2,X_3,X_4$ as in \eqref{X} but with $p_0$ replaced by
$\frac{p_2^2+p_3^2}{4p_1}$. Furthermore, differentiating the function
$p_0:=\frac{p_2^2+p_3^2}{4p_1}$ in $M_4^\circ$ we find $\dot
p_0=X_0|_{M_4^\circ}$ with $X_0$ as in \eqref{X}. This shows that the
restriction to $M_4^\circ$ of the reduced equations of motion are the
restriction to $M_4^\circ$ of the equations $\dot p=X(p)$ with $X$ as
in \eqref{X}. Since the reduced equation of motion is a vector
field in $M_4$, the vector field $X$ is continuous in $\bR5$
and $M_4^\circ$ is dense in $M_4$, by continuity the same is true in
$M_4$.

\vspace{2ex}\noindent
{\bf Acknowledgements.} FF has been partially supported by the
MIUR-PRIN project 20178CJA2B {\it New Frontiers of Celestial
Mechanics: theory and applications}.

\end{document}